\documentclass{article}

\usepackage{microtype}
\usepackage{graphicx}
\usepackage{subfigure}
\usepackage{booktabs} 

\usepackage{hyperref}

\usepackage{algorithmic}
\usepackage[accepted]{mlsys2025}

\usepackage{amsmath,amssymb,amsfonts}

\usepackage[normalem]{ulem}  

\usepackage{cleveref}
\crefname{insight}{Insight}{Insight}

\newcounter{insight}
\newcommand{\boxinsight}[1]{%
  \refstepcounter{insight}%
  \vspace{\smallskipamount}%
  \noindent \simplebox{
    \textbf{\uline{Insight~\theinsight:}} 
    \textit{#1}
  }
}

\usepackage[most]{tcolorbox}
\tcbset{textmarker/.style={%
        enhanced,
        parbox=false,boxrule=0mm,boxsep=0mm,arc=3.5mm,
        outer arc=3.5mm,left=2mm,right=2mm,top=4pt,bottom=3pt,
        toptitle=1mm,bottomtitle=1mm,oversize}}

\newtcolorbox{simplenoteBox}{colback=white, colframe=black, boxrule=0.2mm, arc=1.5mm, auto outer arc, boxsep=0mm, left=2mm, right=2mm, top=1mm, bottom=1mm} 
\newtcolorbox{noteBox}{textmarker,
    colback=gray!8!white}

\newcommand{\simplebox}[1]{\begin{simplenoteBox} #1 \end{simplenoteBox}}

\usepackage{enumitem}

\usepackage{fancyhdr}
\usepackage{appendix}
\usepackage{tcolorbox}
\usepackage{multirow}
\usepackage{booktabs}
\usepackage{listings}
\usepackage{tikz}
\usetikzlibrary{automata, positioning}
\usepackage[table,xcdraw]{xcolor}

\usepackage{multirow}
\usepackage{enumitem}

\definecolor{Highlight}{rgb}{0.92,0.94,1}

\newcommand{\name}[0]{\textit{Sherlock}}

\usepackage[accepted]{mlsys2025}

\mlsystitlerunning{Sherlock: Reliable and Efficient Agentic Workflow Serving with Selective Verification and Speculative Execution}

\begin{document}

\twocolumn[
\mlsystitle{\name: Reliable and Efficient Agentic Workflow Execution }



\mlsyssetsymbol{equal}{*}

\begin{mlsysauthorlist}
\mlsysauthor{Yeonju Ro}{ut,equal}
\mlsysauthor{Haoran Qiu}{azrs}
\mlsysauthor{Íñigo Goiri}{azrs}
\mlsysauthor{Rodrigo Fonseca}{azrs}
\mlsysauthor{Ricardo Bianchini}{azr}
\mlsysauthor{Aditya Akella}{ut}
\mlsysauthor{Zhangyang Wang}{ut}
\mlsysauthor{Mattan Erez}{ut}
\mlsysauthor{Esha Choukse}{azrs}
\end{mlsysauthorlist}

\begin{mlsysaffiliationlist}
\textsuperscript{$\ddagger$}Microsoft Azure Research \hspace{0.5em}
\textsuperscript{$\mathsection$}Microsoft Azure \hspace{0.5em}
\textsuperscript{$\dagger$}The University of Texas at Austin

\end{mlsysaffiliationlist}



\mlsyscorrespondingauthor{Esha Choukse}{esha.choukse@microsoft.com}
\mlsyscorrespondingauthor{Yeonju Ro}{yro@cs.utexas.edu}

\mlsyskeywords{Machine Learning, MLSys}

\vskip 0.3in

\begin{abstract}
With the increasing adoption of large language models (LLM), \textit{ agentic workflows}, which compose multiple LLM calls with tools, retrieval, and reasoning steps, are increasingly replacing traditional applications.
However, such workflows are inherently error-prone: incorrect or partially correct output at one step can propagate or even amplify through subsequent stages, compounding the impact on the final output.
Recent work proposes integrating \textit{verifiers} that validate LLM output or actions, such as self-reflection, debate, or LLM-as-a-judge mechanisms.
Yet, verifying every step introduces significant latency and cost overheads.
In this work, we seek to answer three key questions: which nodes in a workflow are \textbf{\textit{most error-prone}} and thus deserve costly verification, how to select the most \textbf{\textit{appropriate verifier for each node}}, and how to use verification with minimal\textbf{\textit{ impact to latency}}?
Our solution, \name{}, addresses these using counterfactual analysis on agentic workflows to identify error-prone nodes and selectively attaching cost-optimal verifiers only where necessary.
At runtime, \name{} speculatively executes downstream tasks to reduce latency overhead, while verification runs in the background.
If verification fails, execution is rolled back to the last verified output.
Compared to the non-verifying baseline, \name{} delivers an \textbf{18.3\% }accuracy gain on average across benchmarks.
\name{} reduces workflow execution time by up to \textbf{48.7\%} over non-speculative execution and lowers verification cost by \textbf{26.0\%} compared to the Monte Carlo search–based method, demonstrating that principled, fault-aware verification effectively balances efficiency and reliability in agentic workflows.

\end{abstract}

]



\printAffiliationsAndNotice{\mlsysEqualContribution}  

\section{Introduction}
Large language models (LLMs) are increasingly used across domains, from general-purpose tasks such as reasoning~\cite{yao2023react, gou2023tora} and coding~\cite{huang2023agentcoder, zhang2024codeagent} to high-stakes applications like medical diagnosis~\cite{ghezloo2025pathfindermultimodalmultiagentmedical}, scientific research~\cite{aygün2025aihelpscientistswrite}, and data center management~\cite{confucious}.
For complex tasks, a single inference often proves insufficient; instead, these tasks are decomposed into multiple LLM calls (i.e., subtasks) involving tool use, retrieval, and aggregation.
This decomposition results in an \emph{agentic workflow}, naturally modeled as a graph where nodes represent operations (e.g., API call or LLM inference) and edges capture information flow and dependencies.

\textbf{Motivation.}
LLM inference is inherently error-prone and may produce incorrect results due to hallucinations, reasoning flaws, or inadequate context~\cite{yao2023llm, boye2025large}.
In agentic workflows, these errors are particularly problematic: mistakes made in the early nodes propagate down along the edges, amplifying as they go downstream, and contaminate the final output.
To address this, researchers have proposed numerous verification methods, such as self-reflection~\cite{madaan2023self}, debate~\cite{du2023improving}, self-consistency~\cite{wang2023selfconsistencyimproveschainthought}, and LLM-as-a-Judge~\cite{zheng2023judging}, to validate LLM outputs and improve the safety and reliability of LLM inferences.
While effective, each verification incurs additional LLM calls, significantly increasing both latency along the critical path and monetary cost (e.g., up to \textbf{28.9\texttimes{}} and \textbf{53.2\texttimes{}} for instruction-following and coding benchmarks, respectively).
Verifying only the terminal node in a workflow wastes compute resources and misses opportunities to stop error propagation early in the graph, and exhaustive verification across all nodes adds a prohibitively high overhead.

\textbf{Challenges.}
Despite its importance, verification remains poorly understood, with several challenges limiting its deployment.
\uline{First}, identifying vulnerable nodes is difficult—existing frameworks offer little insight into how local errors affect final outcomes, leading to uninformed verifier placement or costly end-to-end tuning~\cite{niu2025flow, zhang2024aflow, hu2024ADAS, sun2021autoflow}.
\uline{Second}, verifier effectiveness and cost vary across tasks and node types, making it hard to balance accuracy and efficiency; expensive verifiers are not always the most cost-effective for a given accuracy.
\uline{Third}, verification can stall execution, as downstream nodes wait for verifiers to finish, creating bottlenecks and latency issues that undermine online responsiveness.

\textbf{Our Work.}
We present \name{}, an agent-serving framework that achieves cost-efficient and reliable execution of agentic workflows through \emph{selective verification} and \emph{speculative execution}.
Unlike static strategies, \name{} adapts to workflow structure and task context through a learning-based design.
It first performs fault-injection–based vulnerability analysis (\S\ref{sec:counterfactual}) to identify error-prone nodes and quantify their influence on final outputs, enabling globally informed verifier placement. The policies learned in this phase are topological, rather than graph-specific, allowing the application of \name{} to dynamically generated agentic workflows.
A learning-based selector (\S\ref{sec:verifier_selector}) then predicts the cost-optimal verifier per node via preference learning that models each verifier’s relative utility {considering performance and cost.}
Finally, a speculative execution runtime (\S\ref{sec:runahead}) overlaps verification and computation to mask verifier latency, adaptively managing rollback to balance latency reduction and re-execution cost.

\name{} enables executing complex and dynamically generated agentic workflows efficiently while maintaining high reliability and controllable verification cost.
In summary, our key contributions are:

\begin{itemize}[leftmargin=*,nosep,topsep=1pt,parsep=6pt]
    \item \textbf{Vulnerability-Guided Verifier Placement}. We perform counterfactual fault injection–based vulnerability analysis to identify error-prone nodes and guide globally informed verifier placement across the workflow. 
    \item \textbf{Cost-Optimal Verifier Selection.} We develop an intelligent, dynamic verifier selector that learns to choose the most cost-effective verifier for a given node, balancing accuracy improvement against verification cost.
    \item \textbf{Speculative Verification Runtime.} We introduce a speculative execution framework that overlaps verification and computation, supported by rollback and recovery mechanisms to ensure correctness under detected errors.
    \item \textbf{End-to-end Evaluation.} We integrate these components into \name{}, a unified system that delivers both high reliability and low latency for complex agentic workflows. Compared to a non-verifying baseline, \name{} delivers an $\textbf{18.3\%}$ accuracy gain on average—best in class compared to verifying baselines, while achieving up to $\textbf{48.7\%}$ execution time reduction and $\textbf{26.0\%}$ cost reduction.
\end{itemize}

\section{Background}

\subsection{Agentic Workflows}
In an agentic workflow, a complex task is decomposed into \emph{subtasks} that collectively form a \textit{workflow}, typically represented as a graph.
Each node in the workflow represents a \emph{subtask} handled by an agent consisting of an LLM call, optionally augmented with tool invocations (e.g., web search or retrieval).
Each edge represents the \emph{output (or history context)} passed from the upstream node to its child node.

\begin{figure}[!t]
  \centering
  \vspace{-0.5em}
  \includegraphics[width=0.8\linewidth]{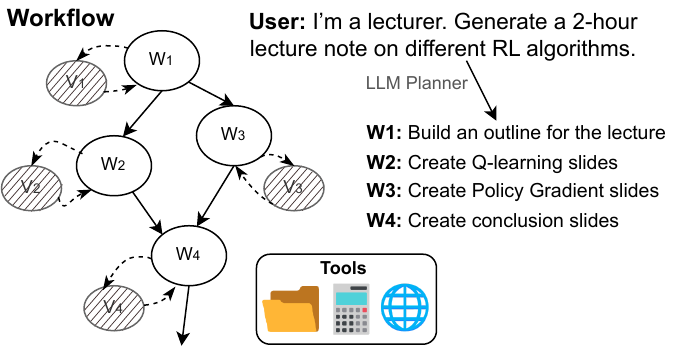}
  \vspace{-1em}
  \caption{\small{
  Example agentic workflow where a user submits a task in natural language and an LLM-based planner generates a workflow composed of multiple subtasks (W1--W4).
  Each node may involve various tools (e.g., web search, file retrieval).
  In this work, we assume adding per-node verifiers (V1--V4).
  }}
  \centering
  \label{fig:workflow}
\end{figure}

An agentic workflow can be statically defined using state-of-the-art agent programming frameworks such as LangGraph~\cite{langgraph} and Agent Framework~\cite{agentframework}.
Recently, \emph{dynamic} workflow generation methods~\cite{sun2021autoflow,niu2025flow,hu2024ADAS,zhang2024aflow} propose using \emph{LLM planners} to constructs the workflow on demand from task descriptions.
\Cref{fig:workflow} shows an example workflow generated from a user task.

\begin{figure}[!t]
  \centering
  \vspace{-0.5em}
  \includegraphics[width=\linewidth]{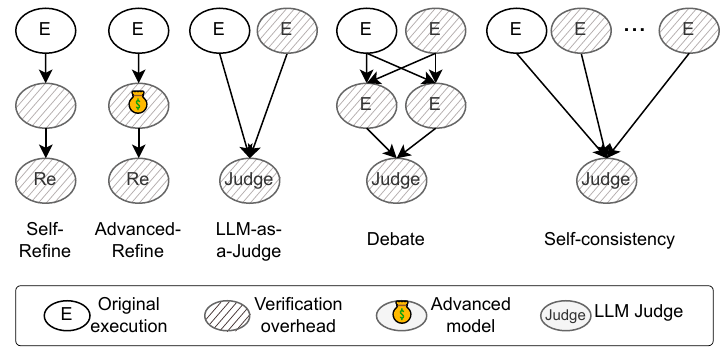}
  \vspace{-2em}
  \caption{\small State-of-the-art LLM verifiers. Grey indicates extra LLM calls from verification, and each dollar emoji indicates an advanced model (more expensive).
  \textit{Judge} indicates a judge LLM. 
  }
  \centering
  \label{fig:verifiers}
\end{figure}

\begin{figure*}[]
  \centering
  \includegraphics[width=\linewidth]{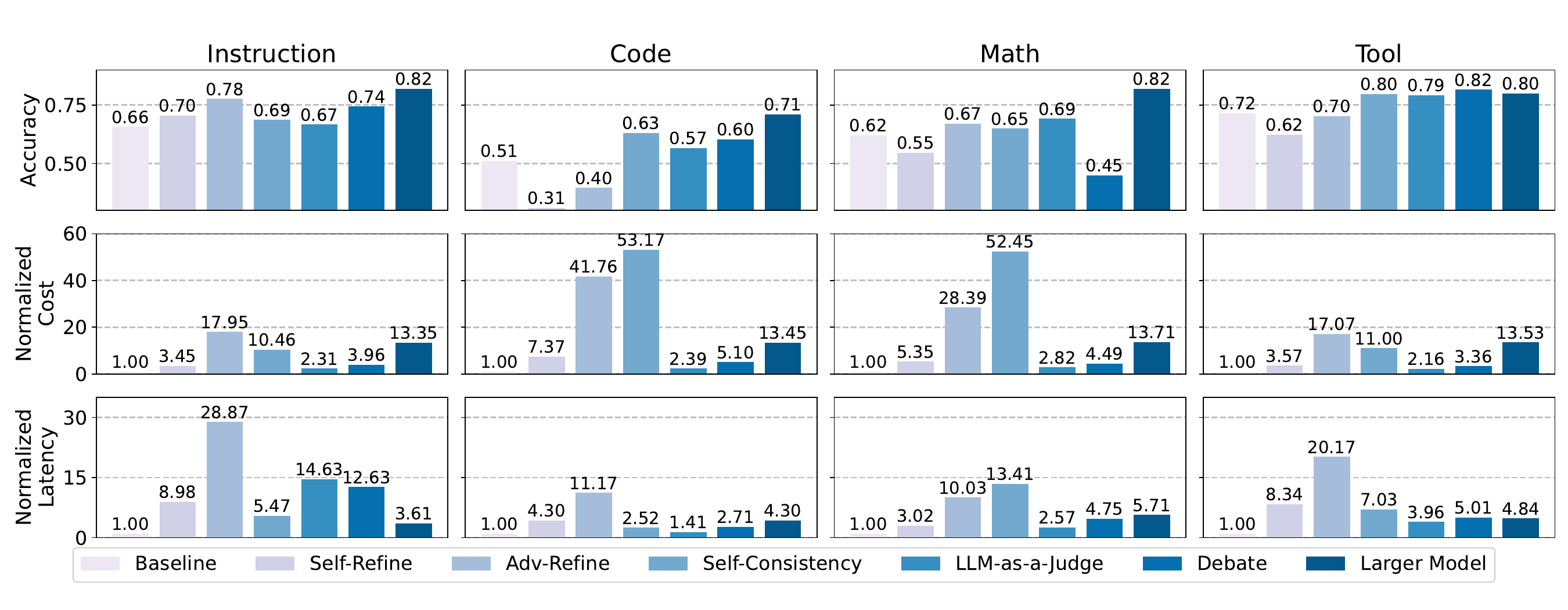}
  \vspace{-3em}
  \caption{\small Verifier Characterization. Comparison of different verifiers' performance across task categories. Latency and cost are normalized to baseline execution latency and cost.
  }
  \centering
  \label{fig:verifier-overhead}
\end{figure*}

\subsection{LLM Verifiers}
LLM outputs and agent actions are error-prone, and may contain hallucinations or logical errors~\cite{cemri2025multi,lin2025llm}, requiring additional \emph{verifier} stages (\Cref{fig:verifiers}).
\emph{Self-refine}~\cite{madaan2023self} recalls the same model to critique and revise its output, while a stronger variant (\emph{Advanced-Refine}) delegates the critique to a larger external model that is more capable but costly.
\emph{Self-consistency} \cite{wang2023selfconsistencyimproveschainthought} instead relies on statistical agreement, sampling multiple answers and trusting the majority.
\emph{LLM-as-a-Judge}~\cite{zheng2023judging} and \emph{Debate} \cite{du2023improving} both introduce an external evaluator: the former compares independent responses, whereas the latter allows the models to argue and refine iteratively before judgment.
Although we focus on these representative verifier types, \name{} can seamlessly integrate new verifier paradigms as they develop.

\section{Verifier Characteristics}
\label{subsec:verifier_characterization}

\subsection{Verifier Overhead} 
While verifiers improve the reliability of LLM outputs, they also add measurable latency and cost.
\Cref{fig:verifier-overhead} quantifies each verifier’s accuracy gain, normalized cost, and latency across task categories\footnote{Details on cost computation and benchmarks in Appx.~\ref{appendix:verifier_overhead}.}.
Verification can increase inference latency and monetary cost by up to \textbf{28.9\texttimes{}} and \textbf{53.2\texttimes{}}, respectively, compared to execution without verification—largely due to additional LLM calls and token usage for feedback generation, judgment, or multi-round reasoning.
These results represent the overhead for a \emph{single verifier} applied to a single output.
In agentic workflows with multiple intermediate nodes, verifying only the final node misses opportunities for early correction and efficient re-execution with verifier feedback, whereas verifying every node compounds overhead through repeated verification steps.

\boxinsight{Verifiers bring significant cost and latency overhead, underscoring the need for principled verifier placement strategies in an agentic workflow.}
\label{insight:overhead}

\begin{figure}[!h]
  \centering
  \vspace{-1em}
  \includegraphics[width=1.05\linewidth, trim=10 0 0 0, clip]{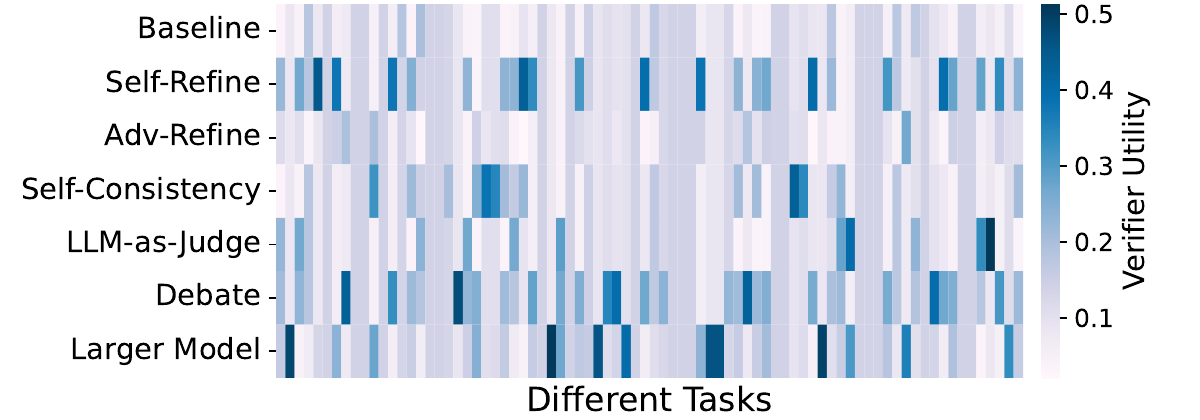}
  \vspace{-1.5em}
  \caption{
  \small Verifier utility by task.
  Utility is computed as $accuracy\_gain - \lambda \cdot cost$, with higher values indicating better cost effectiveness.
  Detailed explanation on verifiers utility in \S\ref{sec:verifier_selector}.
  } 
  \vspace{-1em}
  \centering
  \label{fig:motiv_verifier_utility}
\end{figure}

\subsection{Task-Dependent Accuracy-Cost Tradeoff}

\Cref{fig:verifier-overhead} shows the tradeoff between accuracy and cost across verifiers.
This relationship is \textit{non-monotonic}: higher cost does not necessarily yield better accuracy.
In some cases, excessive or misapplied verification can even \emph{reduce} accuracy, as redundant checks introduce inconsistencies or conflicting judgments.
Because verifier efficacy and cost vary substantially across tasks, certain verifiers provide meaningful accuracy gains, while others incur comparable or higher costs with only marginal improvement.

\Cref{fig:motiv_verifier_utility} shows the utility for each task and verifier.
Tasks within the same category can favor \emph{different} cost-optimal verifiers, revealing fine-grained variability in their effectiveness.
This heterogeneity makes a single global verifier configuration inefficient, underscoring the need for a cost-aware, task-specific verifier selection strategy that dynamically balances accuracy and efficiency at runtime.

\boxinsight{Verifiers have distinct accuracy improvements and cost behaviors, which are highly task dependent.}
\label{insight:tradeoff}

\begin{figure*}[]
  \centering
  \includegraphics[width=\linewidth]{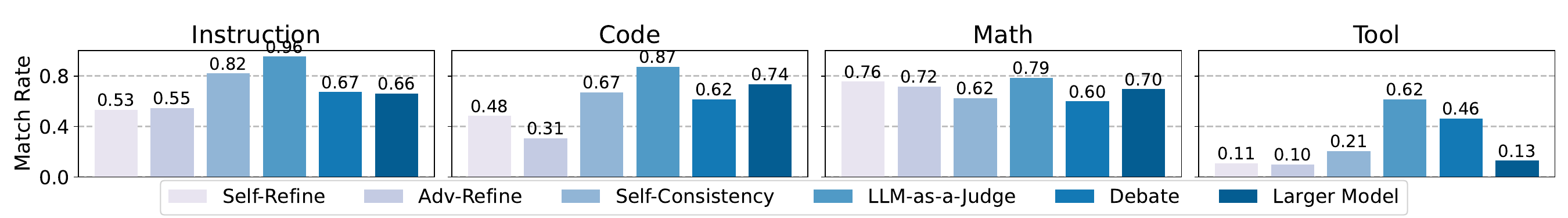}
  \vspace{-1.5em}
  \caption{\small Verified Output Redundancy. Match rate denotes the proportion of verified outputs matching the originals. 
  } 
  \centering
  \label{fig:motiv_verified_output_redundancy}
\end{figure*}

\subsection{Verification Redundancy}
We measure how often a verifier’s revised output matches the original LLM output.
\Cref{fig:motiv_verified_output_redundancy} reports the match rate across task categories.
Most verifiers retain a large portion of original outputs, with match rates commonly above $0.6-0.8$ for Instruction, Code, and Math tasks.
Self-Refine and Adv-Refine show lower rates in Code and Tool tasks, indicating more frequent revisions, while Self-Consistency and LLM-as-a-Judge generally preserve high fidelity.
The Tool category shows the most variation, with some verifiers dropping to 0.10--0.13, underscoring strong task dependence.
Overall, verifications often introduce only minor changes—motivating \emph{speculative execution}, where downstream subtasks proceed in parallel with verification to cut latency without sacrificing output quality.

\boxinsight{Revised output after verification may not necessarily change from its original output.}
\label{insight:redundancy}

\begin{figure}[h]
    \centering
    \includegraphics[width=\linewidth]{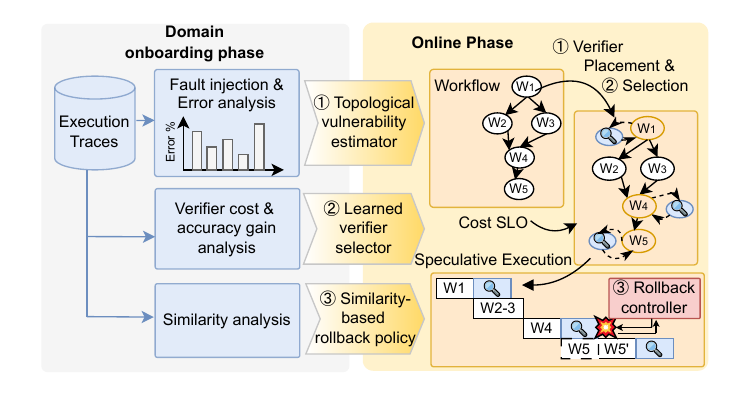}
    \vspace{-2em}
    \caption{
        \small{Overview of the \name{} architecture. }
    }
    \centering
    \label{fig:overview}
\end{figure}

\section{Sherlock Overview}
\label{sec:overview}

Driven by the insights from our characterization (\S\ref{subsec:verifier_characterization}), we design \name{}, a framework that enables reliable and efficient execution of agentic workflows by dynamically 
(1) identifying vulnerable nodes,
(2) selecting and deploying the cost-optimal verifier for those nodes, and
(3) enabling speculative execution to further reduce the performance overhead imposed by verifiers.
To achieve this on dynamically generated agentic workflows, \name{} only requires on-boarding for new \textit{domains}, rather than for each new workflow.

\textbf{Online Phase.}
As depicted in Figure~\ref{fig:overview}, during the online phase of \name{}, a learned topological vulnerability estimator (\S\ref{sec:counterfactual}) is used to decide the priority order of nodes for attaching verifiers, and a learned verifier selector (\S\ref{sec:verifier_selector}) is used to select the cost-optimal verifiers for each of those nodes. All of this is done to maximize the accuracy while meeting the cost budget SLO from the user.
To minimize latency, \name{} employs \textbf{speculative execution} (\S\ref{sec:runahead}), a fast-path strategy that proceeds to subsequent nodes without waiting for verifier results.
If a verifier later revises an output, the similarity between revised output and initial output is computed.
If the similarity is lower than threshold, \textbf{selective rollback} of affected nodes is performed, trading off recomputation for reduced end-to-end latency.
The aggressiveness of speculation and rollback is tunable, allowing adaptation to different reliability–performance trade-offs.

\textbf{Domain On-boarding Phase.}
There are three learned parts in \name{}: a topological vulnerability estimator, a verifier selector, and a similarity threshold to decide whether to rollback.
For a new domain, \name{} needs example workflows with representative execution traces that include the prompts, outputs generated per execution node, and the ground truth for the final output.
\name{} can then analyze these traces to characterize node-level fault patterns and verifier effectiveness.
From these observations, a topology-aware \textbf{verifier placement policy} (\S\ref{sec:counterfactual}) is derived, that prioritizes vulnerable nodes to verify within a given cost budget.
In parallel, a \textbf{verifier selector} (\S\ref{sec:verifier_selector}) is trained, that learns to choose the most cost-efficient verifier for each task prompt and context.
The selector is trained using Group Relative Policy
Optimisation (GRPO)~\cite{shao2024deepseekmath} on preference data defined by the trade-off between accuracy gain and verification cost. 
Furthermore, \name{} quantifies the similarity between generated outputs and ground-truth execution traces, empirically establishing per-metric thresholds to decide when two answers can be considered equivalent.

\name{} distills structural and empirical priors from on-boarding into runtime policies that unify \emph{offline} learning and \emph{online} control, achieving Pareto-optimal trade-offs among accuracy, latency, and cost.

\section{Identifying Error-prone Nodes}
\label{sec:counterfactual}

Given the substantial overhead of verifier execution (\Cref{insight:overhead}), a natural question arises:
\textit{which nodes in a workflow deserve such costly verification?}
To answer that question, we need to understand the fault propagation patterns of workflows.
In this section, we define a fault model, introduce a fault injection method that emulates real-world agent failure modes to derive each node's vulnerability, and finally, propose a topological vulnerability estimator that allows dynamic application to new workflows.

\subsection{Fault Model for Agents}
\label{subsec:fault_model}
Accurate vulnerability estimation requires injected faults to closely replicate realistic failure modes and their empirical occurrence frequencies.
Each node in an agentic workflow can be abstracted as a composition of four elements:
(1) the context received from upstream nodes,
(2) the node’s objective or instruction prompt,
(3) the agent executor (e.g., LLM generation or tool invocation), and
(4) the node’s output.
We apply fault injection to (1), (2), and (4), to understand the importance of each executor node, leading to three primary classes of failure modes (\Cref{tab:fault_mapping}):
\begin{itemize}[nosep,topsep=0pt,parsep=0pt]
\item \textit{Behavioral deviation using prompt replacement}:
simulates behavior deviation by modifying the original task directive.
\item \textit{Context-loss}:
simulates context-loss by removing full or partial conversation history from upstream.
\item \textit{Execution faults using output replacement}:
simulates execution faults by replacing outputs with faulty or inconsistent ones.
\end{itemize}

\begin{table*}[!h]
    \centering
    \small
    \resizebox{1.0\textwidth}{!}{
    \begin{tabular}{lp{7cm}lp{6cm}l}
    \toprule
    \textbf{Failure Mode} & \textbf{Example} & \textbf{Injected Fault} & \textbf{Injected Fault Detail} & \textbf{Frequency} \\
    \midrule
    Behavioral deviation & 
    Misinterpretation of instruction or assigned role & Prompt replacement & Modifying the original task directive & 28.63\% \\ \midrule 
    Context-loss & Loss of full or partial conversational history; missing inter-agent information & Context dropping & Removing full or partial conversation history from upstream & 18.68\% \\ \midrule 
    Execution faults & Erroneous reasoning; task derailment; incoherent reasoning–action sequence; runtime failure (e.g., syntax/formatting errors) & Output replacement & Replacing outputs with faulty or inconsistent ones & 52.69\% \\ 
    \bottomrule
    \end{tabular}}
    \vspace{-1em}
    \caption{Mapping between failure modes, representative examples, corresponding injected faults, and their empirical frequencies.}
    \label{tab:fault_mapping}
\end{table*}

\Cref{tab:fault_mapping} also presents the failure occurrence rates derived from prior work \cite{cemri2025multi}, which analyzed large-scale execution traces of multi-agent workflows to quantify how often each failure class occurs in practice.
We adopt their empirical distribution, after mapping their scenarios to our failure modes, to ensure our injected faults reflect realistic operational conditions\footnote{Frequencies are rescaled after excluding control-flow-related failures (e.g., step repetition, termination control), as well as verification-related failures, which we address in this paper.}.

\begin{figure}[!h]
  \centering
  \includegraphics[width=\linewidth]{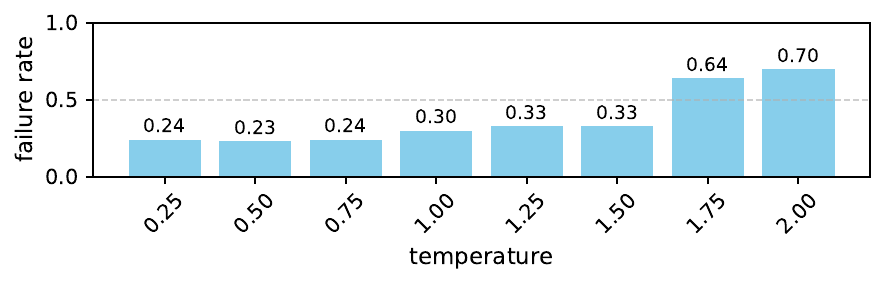}
  \vspace{-2.5em}
  \caption{\small Failure rate by sampling configuration changes.}
  \label{fig:failure_by_temp}
\end{figure}

\subsection{Fault Injection in Workflows}
\label{subsec:fault_injection}
The fault injection begins by executing the pipeline under normal conditions to obtain the baseline output $y$.
For any selected node $n$, we sample a fault from the fault model and generate a counterfactual outputs $\{o'_n\}$ from that node.
Then the downstream computations are re-executed to yield an alternative final result $y'$.
We quantify the deviation $\Delta(y, y')$ for each fault using final accuracy metric, capturing how much the final outcome shifts due to the injected fault.
Finally, we aggregate these deviations into an overall sensitivity score, defined as:
\[
\text{vulnerability\_estimate}(n) = \mathbb{E}_{\text{faults}}[\Delta].
\]
This captures the expected degradation in pipeline correctness if node $n$ were to experience errors, and serves as a basis for guiding verifier placement.

However, a unique confounding factor in LLMs is the stochasticity introduced by sampling configurations (e.g., temperature, Top-$p$, Top-$k$)~\cite{troshin2025control}, which directly affects the observed failure rate (\Cref{fig:failure_by_temp}).
To eliminate this source of randomness, we fix the temperature to 0 during fault injection experiments.

\begin{figure}[!h]
  \centering
  \vspace{-0.5em}
  \includegraphics[width=\linewidth]{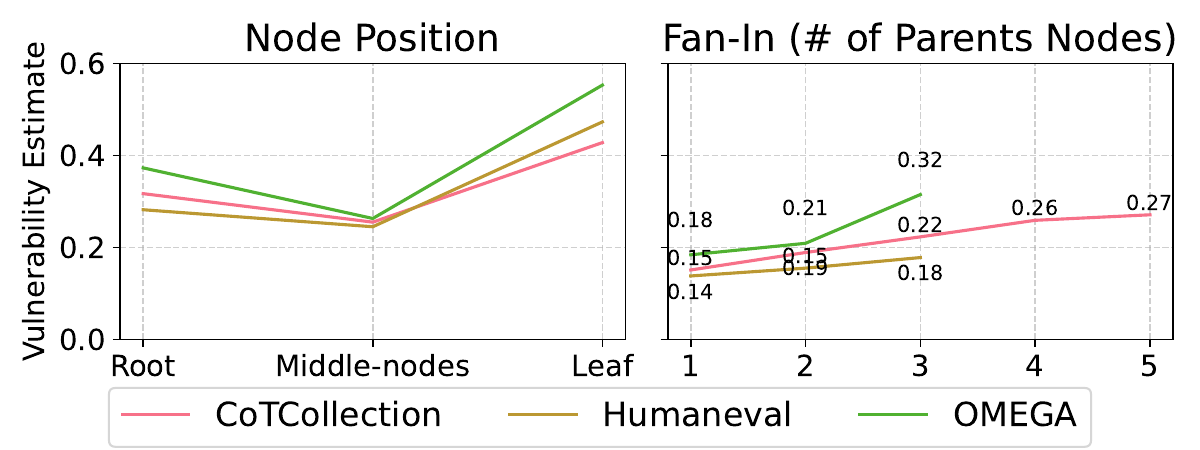}
  \vspace{-1.5em}
  \caption{\small Error distribution by node position and parent count.
  } 
  \centering
  \label{fig:error_distribution}
\end{figure}

\subsection{Topological Error Distribution}
\label{subsec:error_dist}

Based on our fault model, we systematically inject faults into each node of the workflows generated by the LLM planner~\cite{niu2025flow} while running a diverse set of benchmarks:
\textit{CoTCollection}~\cite{kim2023cot}, \textit{OMEGA}~\cite{sun2025omega} and \textit{HumanEval}~\cite{humaneval}, resulting in 100+ different graphs and 15K+ execution traces. 
\Cref{fig:error_distribution} summarizes the aggregated error distribution across all experiments, highlighting how structural and positional properties of workflow nodes affect error propagation, leading to three key observations:

\textbf{Position-wise Sensitivity.}
Terminal nodes are the most vulnerable to faults, followed by initial nodes, while intermediate nodes contribute least to end-to-end failure propagation.
This occurs because downstream nodes can often correct intermediate errors, whereas terminal nodes lack recovery paths.
Initial nodes are an exception as early misinterpretations of the instruction can cascade through the workflow.

\textbf{Fan-in Degree.}
Fan-in degree (the number of incoming dependencies) shows a strong positive correlation with node vulnerability: nodes that aggregate multiple inputs are more likely to amplify or propagate upstream errors.

\textbf{Fan-out Degree.}
We observe little to no correlation between fan-out (the number of child nodes) and the overall error magnitude (therefore not shown in the figure), suggesting that branching alone does not amplify vulnerability.

\subsection{Derived Topology-based Policy}
\label{subsec:verifier_placement}

Motivated by the observed error distribution in the domains of our tested benchmarks, we design a simple yet effective heuristic to guide verifier placement under a cost budget for verification.
Our policy prioritizes nodes that contribute most to final output corruption (\Cref{alg:verifier_heuristic}). 
First, terminal nodes are always selected, as they directly determine the correctness of the final result. 
Next, initial nodes are prioritized to detect early-stage faults that propagate broadly. 
For intermediate nodes, higher fan-in leads to higher priority. Based on this priority, we decide which nodes to verify within the available budget by focusing on the most vulnerable nodes. 

Although we observe that this policy is very robust, and generalizes well across the domains our benchmarks represent, the policy can be learned for each new domain. Furthermore, it can be extended to include more features like model-type, or node functionality, as per the domain's needs. However, it is important to keep it independent of any workflow-specific features, to allow dynamic agentic workflows in runtime.

\begin{algorithm}[h]
    \small
    \begin{flushleft}
    \caption{Topology-driven Verifier Placement}
    \label{alg:verifier_heuristic}
    \begin{algorithmic}[1]
    \REQUIRE Workflow graph $G = (V, E)$, verifier budget $k$
    \STATE $order \gets [\,]$
    \STATE $order.\text{append}({terminal\_node})$
    \STATE $order.\text{append}({initial\_node})$
    \STATE $intermediates \gets \text{sort\_by\_fanin}({intermediate\_nodes})$
    \STATE $order.\text{extend}(intermediates)$
    \STATE $selected \gets order[:k]$ 
    \STATE \textbf{return} $selected$
    \end{algorithmic}
    \end{flushleft}
\end{algorithm}

\section{Cost-Optimal Verifier Selection}
\label{sec:verifier_selector}

After identifying the error-prone nodes to add verifiers, the next step is to select which verifier to attach at each node.
Verifier behavior varies significantly across tasks (\Cref{fig:motiv_verifier_utility}), and their accuracy--cost relationship is highly non-linear:
higher cost does not necessarily translate to higher accuracy (\Cref{insight:tradeoff}).
This task-dependent variability makes a single global configuration or rule-based verifier selection (e.g., building a lookup table) inefficient, motivating a learning-based approach that adapts to node- and task-specific dynamics.
We formalize the verifier selection problem and propose a runtime, learning-based method for dynamic workflows.

\subsection{Verifier Selection Problem}
\label{subsec:problem_formulation}
Given a node to verify, we formulate the verifier selection as a preference learning problem~\cite{chu2005preference}, aiming to learn a policy that captures task-specific preferences among candidate verifiers.
For a set of verifiers $\mathcal{V} = \{v_1, v_2, \dots, v_N\}$ and their observed accuracy-cost pairs $(P(v_i, \tau), C(v_i, \tau))$ on prompts $\tau$ sampled from the dataset~$\mathcal{D}$, we define a preference score for each verifier as
\begin{equation}
U(v_i, \tau) = P(v_i, \tau) - \lambda \, C(v_i, \tau),
\end{equation}
where $\lambda$ is a tunable hyperparameter that controls the trade-off between performance and cost.
$P(v_i, \tau)$ denotes the performance gain (i.e., accuracy improvement) achieved by verifier~$v_i$ on task~$\tau$, and $C(v_i, \tau)$ represents the additional computational cost incurred by using~$v_i$ on task~$\tau$. 

We formulate this as the Lagrangian relaxation of a constrained optimization problem. 
Intuitively, $\lambda$ represents the \textit{willingness to pay}--the marginal cost the system is willing to incur for improved performance. 
\name{} allows users to balance accuracy and cost by tuning the parameter $\lambda$, which serves as a system-level knob to control this trade-off.

\subsection{Preference-Based Policy Optimization}
\label{subsec:training_objective}

We train a policy model $f_\theta(\cdot \mid \tau)$ that, given a task prompt~$\tau \in \mathcal{D}$, outputs a probability distribution over the candidate verifiers $\mathcal{V} = \{v_1, v_2, \dots, v_N\}$. 
Each probability $f_\theta(v_i \mid \tau)$ represents the likelihood of selecting verifier~$v_i$ conditioned on the prompt~$\tau$.
We optimize the policy via Group Relative Policy Optimization (GRPO), which maximizes the expected log-likelihood of verifier selections weighted by their relative advantages:
\begin{equation}
J_{\text{GRPO}}(\theta) = 
\mathbb{E}_{\tau \sim \mathcal{D}}
\Bigg[
\sum_{i=1}^{N}
A(v_i, \tau)
\log f_\theta(v_i \mid \tau)
\Bigg].
\end{equation}
where the advantage term $A(v_i, \tau)$ normalizes each verifier’s preference score within the group:
\begin{equation}
A(v_i, \tau) = U(v_i, \tau) - \frac{1}{N} \sum_{j=1}^{N} U(v_j, \tau).
\end{equation}
In implementation, we minimize the negative objective, $\mathcal{L}_{\text{GRPO}} = -J_{\text{GRPO}}$, to perform gradient descent.

This formulation enables the model to learn verifier preferences that are robust to variations in task utility scales, allowing it to adapt to task-specific trade-offs between accuracy and cost. 
At inference time, given a new task prompt~$\tau$, the Verifier Selector outputs a distribution $f_\theta(v_i \mid \tau)$ over all candidate verifiers, from which the verifier with the highest predicted preference is selected for runtime deployment.

\subsection{Training Data and Model}
\label{subsec:verifier_selector_dataset}

Our training dataset $\mathcal{D}$ consists of task prompts $\tau_i$ (sampled from a diverse set of agentic benchmarks described in Appendix.~\ref{appendix:verifier_overhead}) paired with the observed accuracy and cost of all candidate verifiers $\mathcal{V} = \{v_1, \dots, v_N\}$. 
Each sample is represented as follows.
\[
d_i = \big(\tau_i, \{P(v_j, \tau_i), C(v_j, \tau_i)\}_{j=1}^N \big)
\]
Each prompt $\tau_i$ is encoded into a feature vector $x_i = \phi(\tau_i)$ using the pretrained \texttt{distilbert-base} encoder~\cite{sanh2019distilbert}.
These representations serve as input features for the policy model $f_\theta(\cdot \mid \tau)$, while the preference scores determine the per-task relative advantages during optimization.
The policy model $f_\theta(\cdot \mid \tau)$ applies linear layers to the encoded features to produce logits over $\mathcal{V}$, which are normalized into a preference distribution that defines the model’s selection policy.

\section{Speculative Execution}
\label{sec:runahead}

Our characterization in \S\ref{subsec:verifier_characterization} shows that verifier-revised outputs often remain semantically consistent with the model’s original outputs (\Cref{insight:redundancy}).
This reveals a key inefficiency in existing workflows:
the system idles during verification even though most outputs are correct.
To mitigate this, we introduce \textit{\textbf{speculative execution}}, which overlaps verification of a node with its downstream nodes' computation to hide verifier latency, while being able to \textit{\textbf{rollback}} when the verifier outcomes diverge.

\begin{figure}[h]
  \centering
  \includegraphics[width=\linewidth]{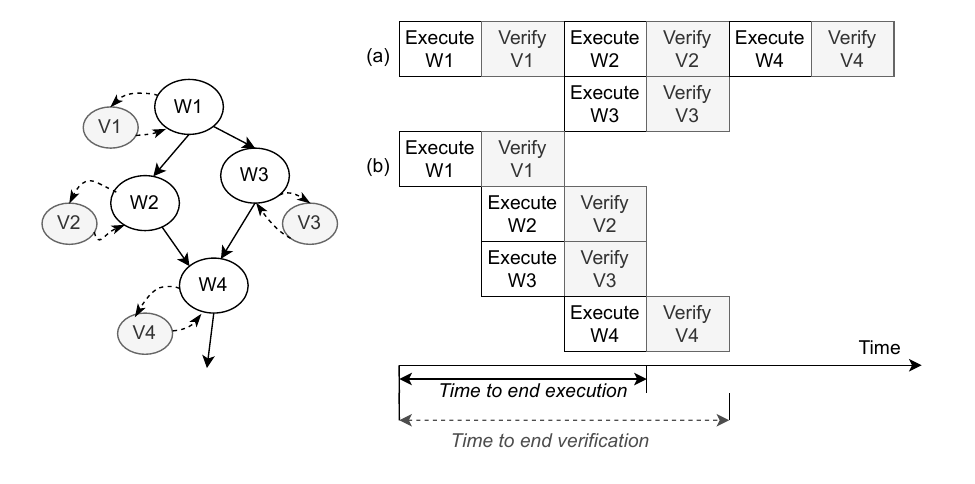}
    \vspace{-2em}
  \caption{\small{An example speculative execution timeline.}}
  \centering
  \label{fig:timeline}
\end{figure}

\Cref{fig:timeline} shows an example of speculative execution.
Once node \texttt{W1} completes, \name{} immediately launches its verifier in the background while concurrently executing child nodes (\texttt{W2}, \texttt{W3}).
If verification later confirms that \texttt{W1}’s output was correct, the speculative results are retained.
Otherwise, they are rolled back to restore correctness.

\begin{figure}[h]
  \centering
  \includegraphics[width=\linewidth]{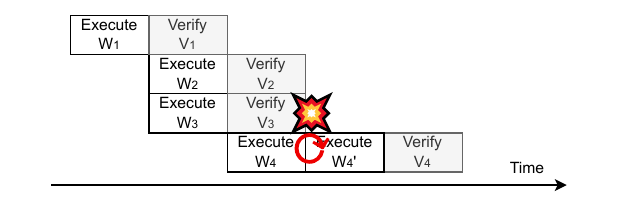}
    \vspace{-2em}
  \caption{\small{Rollback after verification.}}
  \centering
  \label{fig:rollback}
\end{figure}

When verification fails, it indicates that the speculative paths were executed with an incorrect intermediate output.
In this case, \name{} performs a \textbf{rollback} to the failed node, discarding all dependent speculative results (\Cref{fig:rollback}).
The verifier’s corrected output is then propagated forward, and the invalidated nodes are rescheduled for execution.

\subsection{When to Run Ahead: Trade-offs and Cost Model}
\label{subsec:runahead_cost_calculation}

Speculative execution introduces a fundamental \textbf{cost-latency trade-off}.
By speculating deeper along the workflow, \name{} overlaps more downstream computation with verification, reducing total latency but at the risk of wasting more compute when a rollback occurs.
Hence, latency generally decreases at a higher speculative cost.

\textbf{Speculation Depth Bound.}
In \name{}, speculation is bounded by the verifier latency of the current node.
Once the verifier for node~$i$ completes, the system either commits or discards all speculative results based on the verification outcome.
This sets a hard bound on how far the pipeline can safely advance.
We define the set of downstream nodes that can be speculated within this latency window as:

\begin{equation}
\label{eq:n_spec_definition}
N_{\text{spec}} = 
\Bigg\{ 
j \;\Bigg|\;
\sum_{k=i}^{j} \text{lat}_{\text{exec}}^{(k)} 
< 
\text{lat}_{\text{vrf}}^{(i)}
\Bigg\}.
\end{equation}

That is, only nodes in the downstream region whose cumulative execution latency fits within the verifier's latency window are eligible for speculation.

\textbf{Budget-Constrained Speculation.}
Within this bound, users can further tune a \textit{speculation budget}~$B$, which controls how much additional compute overhead is acceptable.
The total speculative cost at node~$i$ must satisfy:
\begin{equation}
\label{eq:budget_constraint}
C_{\text{spec}}(d) \le B,
\end{equation}

The interplay between the verifier latency and cost budget defines the feasible speculation region or depth for each node in the workflow graph.

\textbf{Cost Model.}
To quantify speculative cost, we leverage the \textit{match rate}~$m_i$—the probability that the verifier agrees with the executor output (see Figure~\ref{fig:motiv_verified_output_redundancy}).
Rollback occurs with probability~$(1 - m_i)$, incurring wasted compute across all speculated nodes within~$N_{\text{spec}}$.
The expected speculative cost at node~$i$ is therefore:

\begin{equation}
\label{eq:spec_cost}
C_{spec}^{\text{i}} = (1-m_i) \cdot\sum_{j \in N_{\text{spec}}} 
\left(
C_{\text{exec}}^{j} + C_{\text{vrf}}^{j} 
\right) 
\end{equation}

This formulation ties speculation depth, verifier latency, and budget into a unified model, allowing \name{} to reason about when speculation is beneficial and how aggressively to execute it.

\textbf{Parallel Downstream Execution.}
The previous equations assume sequential downstream execution.
In practice, nodes at the same depth may execute in parallel.
We therefore refine the latency constraint as follows:
\begin{equation}
\label{eq:latency_window_constraint_graph}
\sum_{l=1}^{d} 
\max_{j \in \mathcal{D}_l} 
\text{lat}_{\text{exec}}^{(j)} 
<
\text{lat}_{\text{vrf}}^{(i)},
\end{equation}
where $\mathcal{D}l$ is the set of nodes at depth $l$ downstream of node $i$, and $lat{\text{exec}}^{(j)}$ denotes the execution latency of node $j$.
The summation proceeds over depth levels up to $d$.
Accordingly, the set of eligible speculative nodes is:
\begin{equation}
\label{eq:n_spec_definition_graph}
N_{\text{spec}} =
\Bigg\{
j \;\Bigg|\;
\sum_{l=1}^{\text{depth}(j)}
\max_{k \in \mathcal{D}_l}
\text{lat}_{\text{exec}}^{(k)} 
<
\text{lat}_{\text{vrf}}^{(i)}
\Bigg\}.
\end{equation}

\subsection{Selective Rollback}
\label{subsec:rollback}

When verification revises an output, \name{} determines whether the speculative results can still be retained.
If the verifier’s revision is semantically equivalent to the original output, rollback is unnecessary.
Otherwise, dependent nodes are reverted and re-executed.

To make this decision efficiently, \name{} defines task-specific similarity metrics that quantify output equivalence.
Because LLM responses are often verbose or stylistically diverse, simple string matching is too rigid to capture nuanced similarity~\cite{bulian2022tomayto}.
While LLM-based evaluators can assess semantic similarity more accurately~\cite{adlakha2024evaluating}, they require additional inference calls and introduce substantial latency (i.e., each LLM call takes 2–3 seconds on average, whereas these metrics run in under 0.05 seconds).
To avoid this overhead, \name{} uses lightweight similarity metrics widely adopted in natural language evaluation~\cite{song2024revisiting, niwattanakul2013using, lin2004rouge, papineni2002bleu, makhoul1999performance}.

In the offline analysis (Appx.~\ref{sec:similarity_metric}), we evaluate each metric’s alignment with ground-truth answer equivalence using Spearman correlation and AUC (\Cref{tab:similarity}). 
ROUGE-L achieves the highest consistency for instruction-following and tool-use tasks (\(\rho \approx 0.55\), AUC \(\approx 0.85\)), while all metrics collapse to random performance for code and math (AUC \(\approx 0.5\)). 
Accordingly, at runtime, \name{} retains speculative results when ROUGE-L exceeds a threshold for instruction-following and tool-use tasks, but conservatively defaults to full rollback for code and math tasks.

\section{Evaluation}

\subsection{Setup and Evaluation Methodology}
\textbf{Setup and Models.}
Our experiments run on a server with 8$\times{}$NVIDIA A100 GPUs (80 GB each).
We use vLLM~\cite{kwon2023efficient} to serve the models.
For the base executor, we use meta-llama/Llama-3.1-8B-Instruct~\cite{grattafiori2024llama3herdmodels} on 2 GPUs.
Advanced-Refine verifier uses meta-llama/Llama-3.3-70B-Instruct~\cite{grattafiori2024llama3herdmodels} running on 4 GPUs.
The secondary executor for LLM-as-a-Judge and Debate verifier uses Qwen/Qwen2.5-7B-Instruct~\cite{bai2023qwen} on 1 GPU.
We use AtlaAI/Selene-1-Mini-Llama-3.1-8B~\cite{alexandru2025atla} on 1 GPU as the judge model in LLM-as-a-Judge verifier.

\begin{table}[t]
    \centering
    \footnotesize
    
    \begin{tabular}{llcc}
    \toprule
    \textbf{Component} & \textbf{Model} & \textbf{Size} & \textbf{GPUs} \\
    \midrule
    LLM-as-a-Judge  & Qwen2.5-Instruct        &  7B & 1 \\
    Debate          & Qwen2.5-Instruct        &  7B & 1 \\
    Advanced-Refine & Llama-3.3-Instruct      & 70B & 4 \\
    Judge Model     & Selene-1-Mini-Llama-3.1 &  8B & 1 \\
    \bottomrule
    \end{tabular}
    \vspace{-1em}
    \caption{LLM configurations used for different verifiers. Models of LLM-as-a-Judge and Debate indicate a secondary executor. The primary executor is Llama-3.1-Instruct-8B with 2 GPUs.
    }
    \vspace{0em}
    \label{tab:llm-configs}
\end{table}

\begin{figure*}[]
  \centering
  \includegraphics[width=\linewidth]{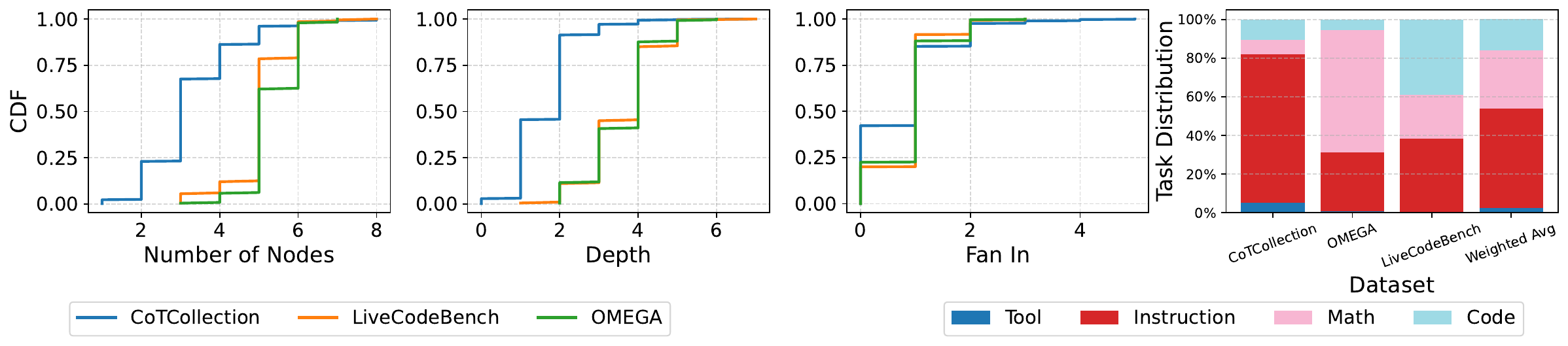}
  \vspace{-1.5em}
  \caption{\small Benchmark characterization for topological properties of the node. 
  } 
  \vspace{-1em}
  \centering
  \label{fig:eval_benchmark_characterization}
\end{figure*}

\textbf{Benchmarks.}
We evaluate \name{} on representative agent benchmarks including CoTCollection~\cite{kim2023cot}, OMEGA~\cite{sun2025omega}, and LiveCodeBench~\cite{jain2024livecodebench}.
CoTCollection consists of instruction-following tasks that demand multi-step reasoning and planning capabilities. OMEGA is a math benchmark, and we specifically use its compositional subset, which requires integrating multiple reasoning skills learned in isolation. LiveCodeBench includes diverse code-related tasks such as code generation, code execution, and test output prediction.
Each benchmark has a mix of instruction-following, coding, math, and tool-calling subtasks as shown in Figure~\ref{fig:eval_benchmark_characterization} (right).

\textbf{Workflow Generation.}
To generate a workflow, we use a state-of-the-art LLM planner, Flow~\cite{niu2025flow}, with customized prompts Appx.~\ref{subsec:llm_planner_prompt}. 
Figure~\ref{fig:eval_benchmark_characterization} shows the generated workflow's characteristics per benchmark.

\textbf{Baselines}
For each component, we compare \name{} against the most relevant baselines.
For verifier placement, we evaluate against even and random placement heuristics under the same cost budget.
For verifier selection, we compare with
(1) \emph{static} selections of the same verifier across the board,
(2) \emph{Aflow}, a Monte Carlo search-based approach~\cite{zhang2024aflow} that incrementally expands workflows by adding new nodes such as debate and self-consistency modules during iterative search,
(3) a \emph{tabular} approach that selects the cost-optimal verifier for each task category, and
(4) an \emph{oracle}, that chooses the cheapest verifier that gives the correct answer for each node.

The \emph{tabular} approach classifies each node into four categories:
instruction-following, coding, math, or tool-use.
We train a task classifier using microbenchmark prompts as inputs and their task categories as labels.
We use \texttt{ModernBERT} as a classifier, which achieves near 98\% task classification F1 score.

\textbf{Evaluation Metrics.}
We evaluate \name{} along three dimensions: accuracy, latency, and cost.
For \textit{accuracy}, we adopt the default metric in each benchmark, i.e., percentage points for CoTCollection and OMEGA, and pass@1 rate for LiveCodeBench.
For \textit{latency}, we define two complementary metrics: (1) \uline{\textit{Time to End Execution}} (T\textsubscript{exec}) measures how long it takes to finish all execution in one graph, reflecting the latency of \textit{the fast path} that is achievable by speculative execution.
(2) \uline{\textit{Time to End Verification}} (T\textsubscript{vrf}) measures the total duration until the final verified output is available, representing the latency of \textit{the slow path} that includes all verification stages.
\Cref{fig:timeline} visualizes these two metrics.
Finally, we compute the \textit{verifier cost} according to a cost model (Appx.~\ref{subsec:verification_cost}) since we serve open-source models locally.

\subsection{Verifier Placement}
\begin{figure}[!h]
  \centering
  \includegraphics[width=0.85\linewidth]{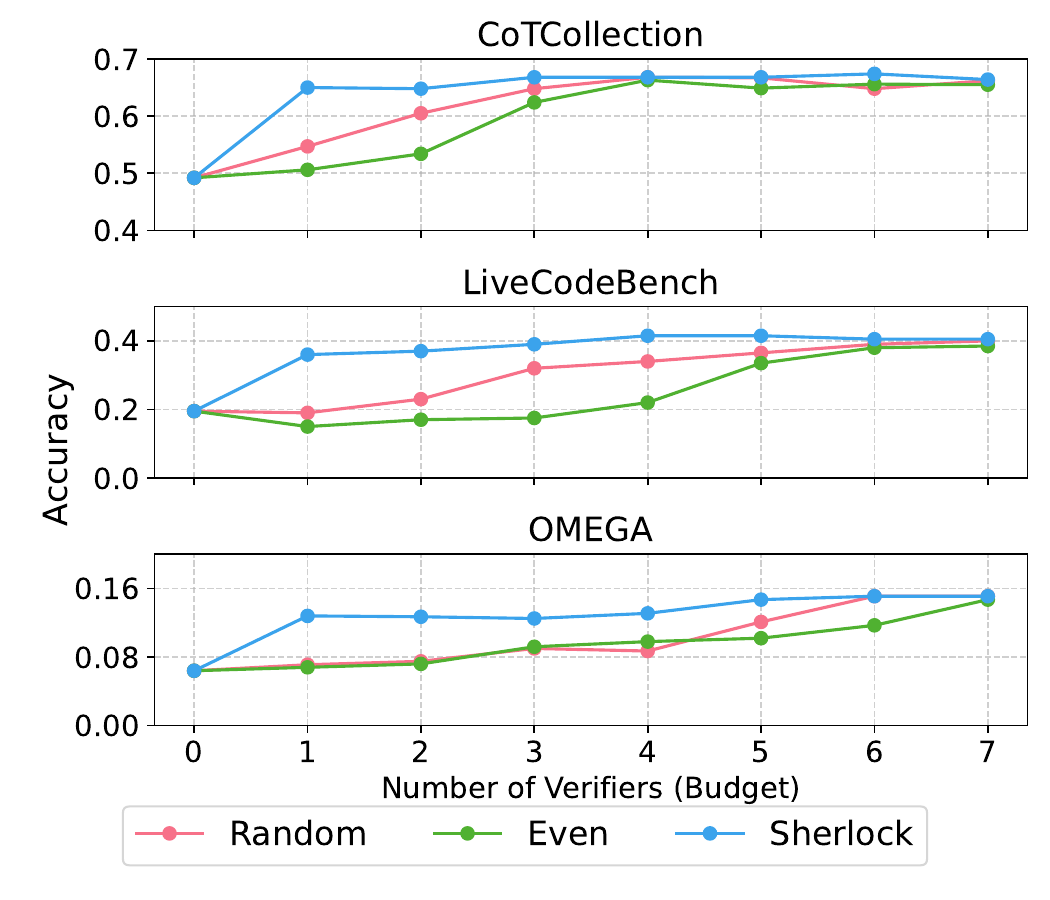}
  \vspace{-1em}
  \caption{\small Accuracy improvement with different placement strategy. \name{} uses the policy described in \Cref{subsec:verifier_placement}.
  }
  \label{fig:eval_placement}
\end{figure}

We evaluate \name{}’s topology-aware verifier placement (\S\ref{subsec:verifier_placement}) by comparing it to random and evenly spaced placements under the same verification budget.
All methods use the same verifier selection, isolating placement as the only variable.
\Cref{fig:eval_placement} shows that \name{} consistently achieves higher final accuracy with less budget.
This demonstrates that \name{} allocates verification resources more effectively than uninformed strategies by leveraging observed error distributions.

\begin{figure}[!h]
  \centering
  \includegraphics[width=\linewidth]{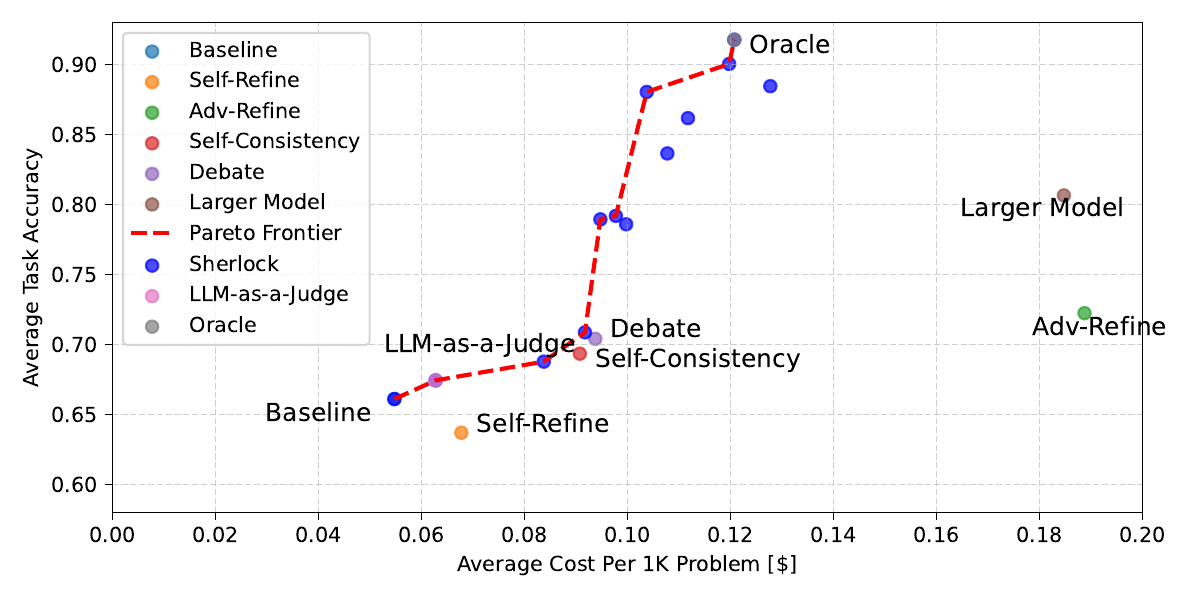}
  \vspace{-2.5em}
  \caption{\small Accuracy-cost trade-off across verifier selections. \name{} (blue dots) lies on the Pareto frontier (red dashed line) and approaches the oracle, achieving high cost efficiency.
  }
  \label{fig:eval_selection_micro}
\end{figure}

\subsection{Verifier Selection}
\Cref{fig:eval_selection_micro} shows the accuracy--cost trade-off achieved by \name{}'s verifier selector (\S\ref{sec:verifier_selector}).
Using the microbenchmark described in Appx.~\ref{appendix:verifier_overhead}, we train the selector and evaluate its accuracy and average cost per task based on the verifier it chooses.
For each problem, we define the oracle verifier as the one that achieves the highest accuracy gain at the lowest cost.
Compared to static verifier assignments, \name{} consistently follows the Pareto frontier.
It achieves higher accuracy for the same cost and approaches the accuracy of the oracle (i.e., the best accuracy that can be achieved with the given set of executors and verifiers).

\begin{figure}[!h]
  \centering
  \includegraphics[width=\linewidth]{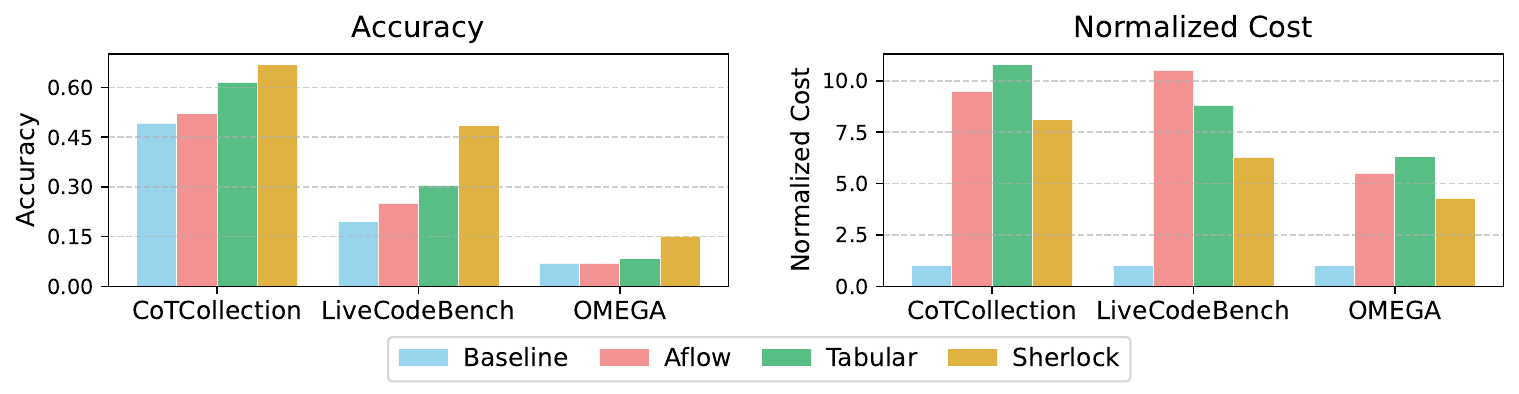}
  \vspace{-2.5em}
  \caption{\small End-to-end comparison with the state-of-the-art approaches in final accuracy and verifier cost.}
  \label{fig:eval_selection_e2e}
\end{figure}

Using the cost-optimal verifier selector checkpoint, we evaluate the end-to-end accuracy of three agentic benchmarks with \name{}.
\Cref{fig:eval_selection_e2e} shows the final accuracy and total verification cost comparison.
Compared to the tabular approach, \name{} consistently achieves higher accuracy at a lower cost.
We also compare \name{} against AFlow~\cite{zhang2024aflow}, a Monte Carlo Tree Search–based framework for discovering agentic workflows.
Across all tasks, \name{} consistently delivers higher accuracy and lower cost, primarily due to its dynamic verifier selector that adapts flexibly to task characteristics.

\subsection{Speculative Execution}
\label{subsec:eval_spec_rollback}

\begin{figure}[!h]
  \centering
  \vspace{-0.5em}
  \includegraphics[width=0.8\linewidth]{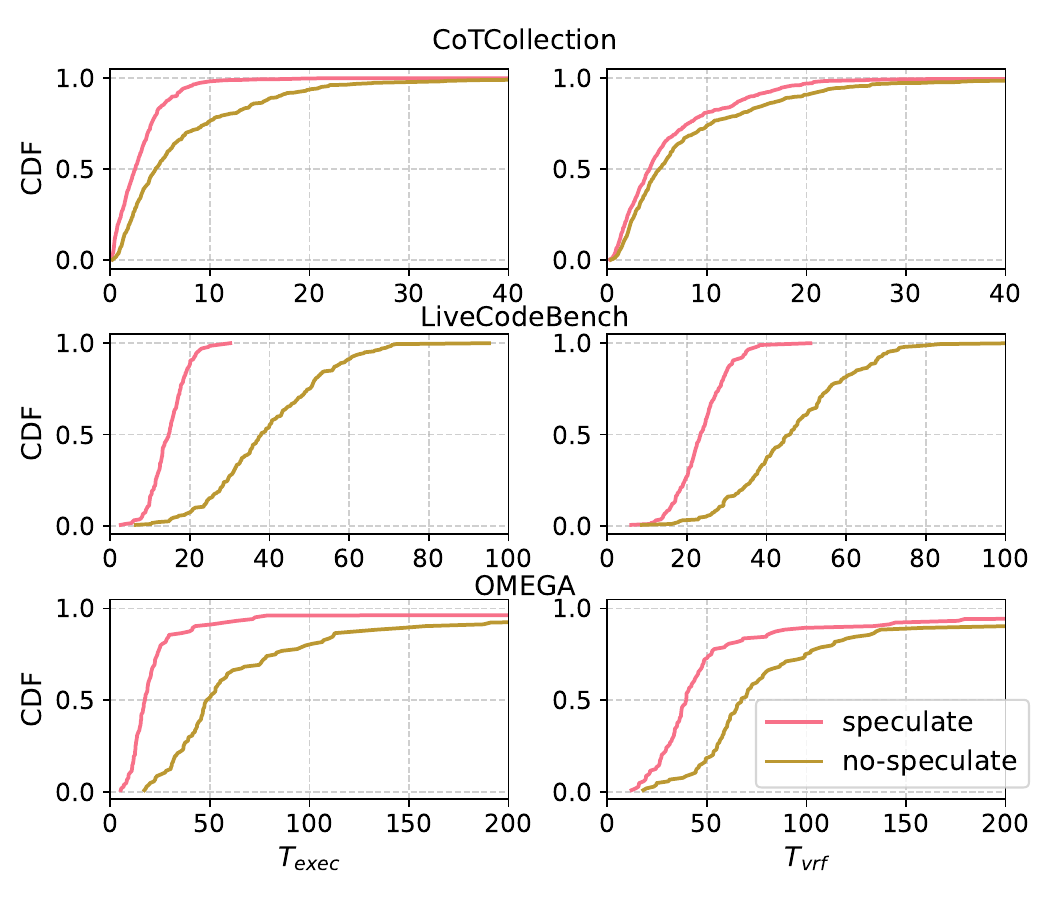}
  \vspace{-1em}
  \caption{\small Comparison of Time to end execution (\(T_{\text{exec}}\)) and Time to end verification (\(T_{\text{vrf}}\)) regarding \name{} speculation.}
  \label{fig:eval_speculative_execution_cdf}
\end{figure}

\Cref{fig:eval_speculative_execution_cdf} shows the CDFs of \name{}’s time to end execution (\(T_{\text{exec}}\)) and time to end verification (\(T_{\text{vrf}}\)) latencies (\S\ref{sec:runahead}). \Cref{tab:latency_improvement} summarizes the latency reductions achieved through speculative execution.
Speculative execution substantially reduces latency across all benchmarks, with the largest gains observed in \textbf{LiveCodeBench}, where mean \(T_{\text{exec}}\) drops by \textbf{62.9\%} and \(T_{\text{vrf}}\) by \textbf{48.7\%}.
These results show that overlapping verification with downstream execution can effectively hide verifier delays, particularly in complex multi-step reasoning tasks.
CoTCollection and OMEGA also show consistent $T_{\text{exec}}$ reductions exceeding \textbf{50\%}, confirming that the benefit generalizes across diverse workflows and latency profiles.

\begin{table}[t!]
\footnotesize
\centering
\resizebox{\linewidth}{!}{
\begin{tabular}{lccccc}
\toprule
                               &                          & Mean   & P50    & P90    & P99    \\ \midrule
\multirow{2}{*}{CoTCollection} & T\textsubscript{exec}    & 51.4\% & 44.8\% & 60.8\% & 67.7\% \\
                               & T\textsubscript{vrf} & 21.9\% & 18.5\% & 24.9\% & 52.0\% \\ \midrule
\multirow{2}{*}{OMEGA}         & T\textsubscript{exec}    & 53.4\% & 64.5\% & 72.7\% & 4.8\%  \\
                               & T\textsubscript{vrf} & 30.6\% & 41.1\% & 34.7\% & 5.5\%  \\ \midrule
\multirow{2}{*}{LiveCodeBench} & T\textsubscript{exec}    &  62.9\%  &  61.3\%  & 65.9\%  & 62.1\%  \\
                               & T\textsubscript{vrf} & 48.7\%  & 49.1\%  & 53.3\%  & 52.6\% \\ \bottomrule
\end{tabular}
}
\vspace{-1em}
\caption{Latency reduction with speculative execution.
}
\vspace{-1em}
\label{tab:latency_improvement}
\end{table}



\section{Related work}

\textbf{Agentic Workflow Generation and Optimization.}
Recent work has explored automating agentic workflow generation to improve response quality through iterative search~\cite{hu2024ADAS, zhang2024aflow}, LLM-based generation~\cite{li2024autoflow, liu2025divideoptimizemergefinegrained, niu2025flow}, and fine-tuning~\cite{wu2025optimas}.
This line of work is complementary to \name{} and can benefit from \name{}'s selective verification and speculative execution to balance accuracy, cost, and latency.
Compared to iterative-search or fine-tuning-based approaches, \name{} employs a lightweight strategy that combines heuristics and learned verifier selection for fast adaptability to online tasks.
Compared to LLM-based workflow generators, \name{} systematically augments generated workflows with its dynamic verification policies.

\textbf{Serving Agentic Workflows.}
Murrakab~\cite{chaudhry2025murakkab} highlights the inefficiency of existing agent-serving systems that treat workflows as opaque sequences of model and tool calls, tightly coupling agent logic with hardware and model choices.
Circinus~\cite{liu2025circinus} introduces an SLO-aware query planner for compound AI workloads, optimizing operator placement and configuration across heterogeneous infrastructure.
Complementary to these, Parrot~\cite{luo2025empiricalstudycatastrophicforgetting}, Autellix~\cite{luo2025autellix}, and AI Metropolis~\cite{xie2024ai} explore scheduling and orchestration for multi-step agentic workflows.
LLMSelector~\cite{chen2025optimizing} further shows that end-to-end performance improves with stronger individual nodes in a workflow.
Speculative Actions~\cite{ye2025speculative} and Conveyor~\cite{xu2024conveyor} enable asynchronous execution of LLM actions and tool calls in the background.
\name{} is the first framework that holistically explores the trade-offs between cost, accuracy, and latency by exploiting speculative execution opportunities with intelligent verifier selection for agentic workflows.

\section{Conclusion}

We presented \name{}, a principled serving framework for agentic workflows that jointly optimizes latency, cost, and accuracy.
\name{} identifies and verifies error-prone nodes through counterfactual analysis and dynamic verifier selection, effectively balancing reliability and efficiency.
To further reduce verification overhead, it employs selective speculative execution and rollback, overlapping verification with downstream computation while controlling rollback cost.
\name{} also exposes user-configurable knobs to flexibly trade off reliability, latency, and cost on demand.
Overall, \name{} delivers up to \textbf{48.7\%} execution latency reduction and \textbf{26.0\%} cost reduction over baselines.
 
\bibliography{example_paper}
\bibliographystyle{mlsys2025}

\clearpage

\appendix
\onecolumn

\section{Cost and Latency Overhead of Verifiers}
\label{appendix:verifier_overhead}

\subsection{Microbenchmarks}
\label{subsec:microbenchmarks}
In this paper, we use following microbenchmarks. 
\begin{itemize}
    \item Instruction-following tasks: HotpotQA~\cite{yang2018hotpotqa}, DROP~\cite{dua2019drop}, Instruction~\cite{hfinstruction}
    \item Math tasks: MATH~\cite{math500}, GSM8k~\cite{gsm8k}
    \item Coding tasks: Humaneval~\cite{humaneval}, MBPP~\cite{austin2021program}
    \item Tool-calling tasks: GTA~\cite{wang2024gta} 
\end{itemize}

\subsection{Verifier Cost Calculation}
\label{subsec:verification_cost}

Each verifier incurs different costs, which can be calculated as the sum of model cost and GPU cost. Model cost refers to the expense of using closed-source APIs—for example, commercial LLMs—whereas open-source models like the LLaMA family~\cite{grattafiori2024llama3herdmodels} and Qwen family~\cite{bai2023qwentechnicalreport} do not incur model costs. However, serving these open-source models requires compute resources, leading to GPU costs.

Model cost typically depends on the number of tokens in the prompt (input) and response (output), with output tokens generally priced higher. GPU cost is determined by the number of GPUs used and the duration of use. Given unit cost, GPU cost can be calculate as follows. 

\begin{equation}
\label{eq:gpu_cost}
\begin{split}
    GPU\_cost= & \frac{unit\_price \times num\_gpus \times num\_tokens }{throughput_{max}\times cluster\_utilization_{avg}}
\end{split}
\end{equation}

In our setup, we deploy open-source models for the executors, advanced feedback model, and judge on an 8-GPU server costing \$13.60 per hour (\textit{unit\_price}). We allocate 2 GPUs to the main executor model, 1 GPU to secondary executor model,  1 GPU to the judge model, and 4 GPUs to the advanced feedback model.

\clearpage

\subsection{Judge LLM and Majority Answer}
For Self-consistency, the majority answer can be obtained either by generation (\texttt{sc-gen}), where a new response is produced from the set of executor outputs, or by selection (\texttt{sc-select}), where an existing executor response closest to the majority is chosen. \texttt{sc-gen} provides higher-quality outputs by smoothing superficial variations (e.g., wording, formatting) but incurs additional generation cost. By contrast, \texttt{sc-select} is more efficient, requiring only a single token to identify the nearest executor, but cannot reconcile divergent responses when no exact overlap exists.

\noindent
The relative effectiveness of \texttt{sc-gen} and \texttt{sc-select} varies across tasks, largely due to the specialization of the Judge LLM. Since the Judge model is fine-tuned specifically for evaluation, such specialization can degrade its general instruction-following ability~\cite{luo2025empiricalstudycatastrophicforgetting, wang2024twostagellmfinetuningspecialization}, often leading it to disregard formatting constraints given in prompts. This negatively impacts tasks requiring strict output structure, such as code generation or tool invocation. Consequently, \texttt{sc-gen} with a general-purpose model (e.g., LLaMA-8B) outperforms \texttt{sc-gen} with the Judge LLM on Tools and Code tasks. In contrast, for \texttt{sc-select}, the Judge LLM proves more effective, yielding superior solutions on Tools and Code tasks and achieving comparable accuracy on QA tasks (Tab~\ref{tab:sc_select}).

\begin{table}[!h]
\centering
\caption{\small{Generation-based vs. Selection-based Majority Answer (sc-gen vs. sc-select)}}
\resizebox{0.5\linewidth}{!}{
\begin{tabular}{l|cc|c|cc}
\toprule
                      & \multicolumn{2}{c|}{QA} & \multicolumn{1}{c|}{Tools} & \multicolumn{2}{c}{Code} \\
                      & Drop       & Hotpot    & GTA                       & Humaneval    & MBPP      \\ \midrule
No-verify             & 59.0\%     & 74.5\%    & 65.8\%                    & 60.8\%       & 44.5\%    \\ \midrule
sc-gen (Judge LLM)    & 58.0\%     & 70.5\%    & 39.7\%                    & 65.2\%       & 49.9\%    \\
sc-gen (llama3-8b)    & 55.0\%     & 68.0\%    & 50.4\%                    & 65.9\%       & 52.0\%    \\ \midrule
\rowcolor{Highlight}
sc-select (Judge LLM) & 62.0\%     & 75.5\%    & \textbf{71.4\%}                    & \textbf{76.2\%}       & \textbf{58.1\% }   \\
\rowcolor{Highlight}
sc-select (llama3-8b) & \textbf{63.0\%}     & \textbf{76.0\%}    & 67.7\%                    & 69.5\%       & 57.4\%   \\ \bottomrule
\end{tabular}
}
\label{tab:sc_select}
\end{table}

\section{Speculative Execution State Machine Diagram }
\label{appendix:state_machine_diagram}

\begin{figure}[ht]
    \centering
    \begin{minipage}[t]{0.5\textwidth}
        \textbf{FSM Definition:}
        \begin{itemize}
            \item \small{States: \( Q = \{\texttt{waiting}, \texttt{running}, \texttt{verifying}, \texttt{completed}, \texttt{failed} \} \)}
            \item \small{Alphabet: \( \Sigma = \{\texttt{run}, \texttt{verify}, \texttt{no-verify}, \texttt{success}, \texttt{fail}, \texttt{rerun} \} \)}
            \item \small{Initial state: \( q_0 = \texttt{waiting} \)}
            \item \small{Accepting state: \( F = \{\texttt{completed} \} \)}
            \item \small{Transition function \( \delta \) is defined as:}
        \end{itemize}

        \vspace{-1.5ex} 
        \[
        \small 
        \begin{array}{|c|c|c|}
            \hline
            \text{Current State} & \text{Input} & \text{Next State} \\
            \hline
            \texttt{waiting} & \texttt{run} & \texttt{running} \\
            \texttt{running} & \texttt{verify} & \texttt{verifying} \\
            \texttt{running} & \texttt{no-verify} & \texttt{completed} \\
            \texttt{verifying} & \texttt{success} & \texttt{completed} \\
            \texttt{verifying} & \texttt{fail} & \texttt{failed} \\
            \texttt{failed} & \texttt{rerun} & \texttt{completed} \\
            \hline
        \end{array}
        \]
    \end{minipage}\hfill
    \begin{minipage}[t]{0.45\textwidth}
        \vspace{7em}
        \textbf{State Diagram:}
        \begin{center} 
        \begin{tikzpicture}[
            shorten >=1pt, 
            node distance=2.5cm, 
            on grid, 
            auto, 
            scale=0.8, 
            transform shape 
        ]
            \node[state, initial] (W) {waiting};
            \node[state] (R) [right=of W] {running};
            \node[state] (V) [right=of R] {verifying};
            \node[state, accepting] (C) [below=of R] {completed};
            \node[state] (F) [below=of V] {failed};

            \path[->]
                (W) edge node {run} (R)
                (R) edge node {verify} (V)
                (R) edge[bend right] node[swap] {no-verify} (C)
                (V) edge node[swap] {success} (C)
                (V) edge node {fail} (F)
                (F) edge[bend right] node[pos=0.55, swap] {rerun} (C);
        \end{tikzpicture}
        \end{center}
    \end{minipage}
    \caption{Finite State Machine definition and diagram}
    \label{fig:fsm-definition}
\end{figure}
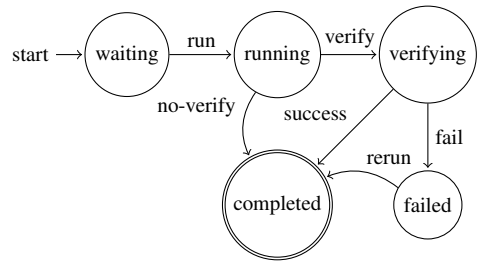

Internally, \name\ manages workflow execution using a node-state machine. 
Each node begins in the \texttt{waiting} state, transitions to \texttt{running} during computation, and then to either \texttt{verifying} or \texttt{completed}, depending on whether verification is required.
While a node is \texttt{verifying}, its child nodes may begin execution speculatively.
If verification succeeds, the node transitions to \texttt{completed}; if verification fails, it transitions to \texttt{failed}, triggering rollback (\S\ref{subsec:rollback}). 
The full state transition diagram is shown in Figure~\ref{fig:fsm-definition} .

\clearpage

\section{Lightweight Similarity Metrics for Practical Rollback}
\label{sec:similarity_metric}

\begin{table}[!h]
\centering
\small
\setlength{\tabcolsep}{4pt}
\renewcommand{\arraystretch}{1.2}
\begin{tabular}{lcc|cc|cc|cc}
\toprule
\multirow{2}{*}{\textbf{Metric}} & 
\multicolumn{2}{c|}{\textbf{Instruction}} & 
\multicolumn{2}{c|}{\textbf{Tool}} & 
\multicolumn{2}{c|}{\textbf{Code}} & 
\multicolumn{2}{c}{\textbf{Math}} \\
\cmidrule(lr){2-9}
 & \textbf{Spearman} & \textbf{AUC} & 
   \textbf{Spearman} & \textbf{AUC} &
   \textbf{Spearman} & \textbf{AUC} &
   \textbf{Spearman} & \textbf{AUC} \\
\midrule
Cosine Similarity   & 0.314 & 0.687 & 0.368 & 0.717 & 0.028 & 0.516 & $-$0.324 & 0.262 \\
Jaccard Similarity  & 0.411 & 0.740 & 0.272 & 0.620 & 0.134 & 0.577 & $-$0.436 & 0.237 \\
\rowcolor{gray!10}
RougeL   & \textbf{0.556} & \textbf{0.828} & \textbf{0.668} & \textbf{0.877} & 0.099 & 0.557 & $-$0.205 & 0.349 \\
BLEU     & 0.381 & 0.723 & 0.267 & 0.618 & 0.057 & 0.533 & $-$0.449 & 0.230 \\
AST Similarity & 0.200 & 0.589 & 0.427 & 0.712 & -- & 0.500 & -- & 0.500 \\
\bottomrule
\end{tabular}
\caption{Spearman correlation and AUC of similarity metrics across task categories. 
Higher values indicate stronger alignment with ground-truth answer equivalence. 
ROUGE performs best for instruction and tool tasks, while all metrics fail for code and math (AUC $\approx$ 0.5).}
\label{tab:similarity}
\end{table}

To determine whether a verifier’s revision is different enought to invalidate speculative executions on initial answer, we seek lightweight similarity metrics that can replace expensive LLM-based judges. Prior work commonly uses LLM evaluators~\cite{adlakha2024evaluating,bulian2022tomayto}, but invoking them along the critical path adds substantial latency and cost. 
Instead, we examine a set of interpretable alternatives—cosine similarity, Jaccard similarity~\cite{niwattanakul2013using}, ROUGE-L~\cite{lin2004rouge}, BLEU~\cite{papineni2002bleu}, and AST similarity~\cite{song2024revisiting}.

We assess each metric’s alignment with the ground-truth equivalence labels (collected once using an LLM judge for calibration) via Spearman correlation and area under the ROC curve (AUC), as summarized in \Cref{tab:similarity}. 
For instruction-following and tool-use tasks, ROUGE exhibits the strongest association with the ground-truth labels (\(\rho \approx 0.56-0.77\), AUC \(\approx 0.83\)–\(0.88\)), indicating that lexical overlap reliably captures answer equivalence in these natural-language settings. 
In contrast, for code-generation and math-reasoning tasks, all metrics collapse to random or even negative correlation (\(\rho < 0.2\), AUC \(\approx 0.5\)), confirming that lightweight metrics do not sufficiently capture the semantic equivalence.

\begin{figure}[!h]
  \centering
  \includegraphics[width=\linewidth]{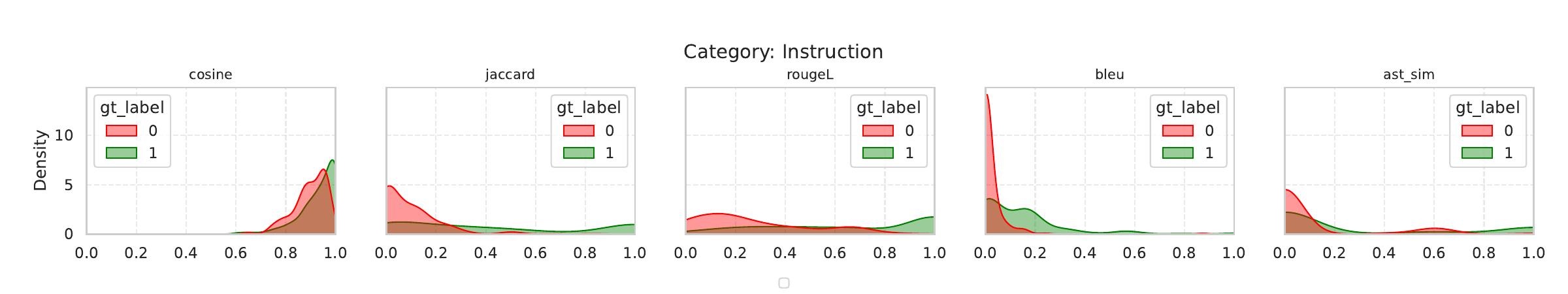}
  \includegraphics[width=\linewidth]{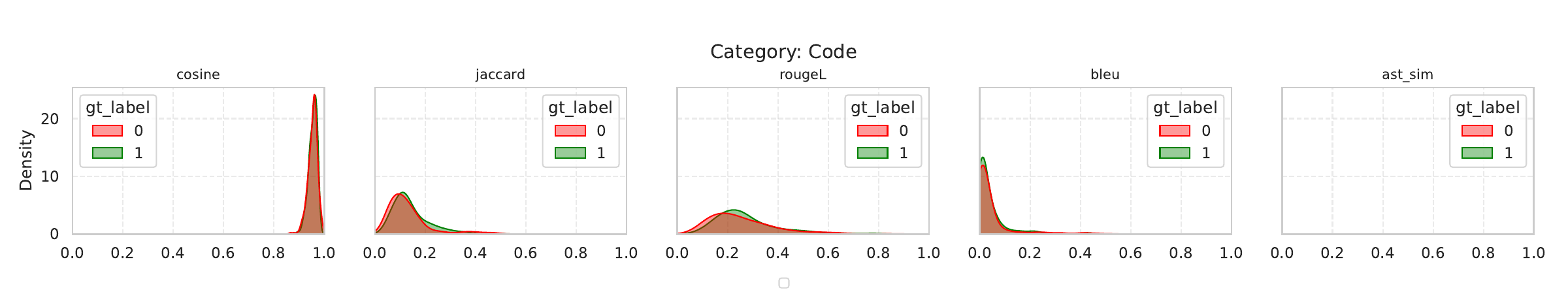}
  \includegraphics[width=\linewidth]{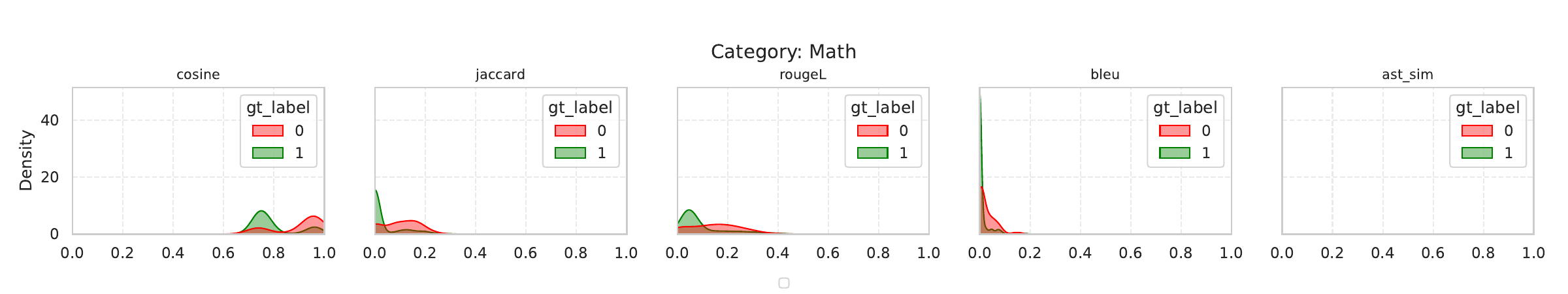}
  \includegraphics[width=\linewidth] 
  {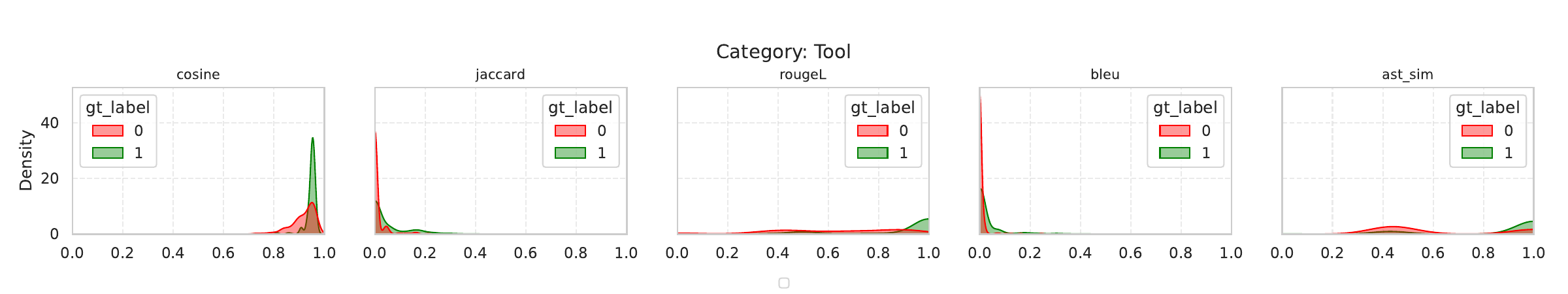}
  \caption{\small Kernel Density Estimation (KDE) plots of similarity metrics across task categories. Each subplot compares the distributions of similarity scores for match with ground truth ($gt\_label=1$) and not matching with ground truth ($gt\_label=0$) outputs.}
  \label{fig:appx_similarity_kde}
\end{figure}

To further examine discriminative behavior, we visualize the kernel density estimation (KDE) of similarity scores for matching (\(gt\_label=1\)) and non-matching (\(gt\_label=0\)) pairs in \Cref{fig:appx_similarity_kde}. 
While instruction and tool-use tasks show clear separation between the two distributions, code and math tasks exhibit substantial overlap—consistent with their near-random AUC values despite superficial shape differences in the KDE plots. 

These findings collectively indicate that lightweight metrics can safely replace LLM judgment for natural-language tasks but fail to capture semantic equivalence in structured or symbolic domains. 
Consequently, we conservatively default to {rollback} for code and math categories, avoiding false equivalence caused by spurious surface-level similarity.

\clearpage

\section{Prompts}
\label{sec:prompts}

\subsection{Scorer Prompt}
\label{subsec:scorer_prompt}

\begin{center}
\begin{tcolorbox}[width=0.95\textwidth]
You are a scorer that tells whether two answers are same. I'll provide Ground truth and Prediction. 
You have to tell that whether the prediction is saying a correct answer same as the ground truth. 
If the prediction is a sentence that include the ground truth, then it is correct answer. 
If the prediction is an equation that ends up with the ground truth, then it is correct answer.

If the answer is correct, output your final verdict by strictly following this format: "[[Correct]]"
If the answer is incorrect, output your final verdict by strictly following this format: "[[Incorrect]]" 

\vspace{1em}

Ground truth: $<$\texttt{GROUND\_TRUTH}$>$

Prediction: $<$\texttt{PREDICTION}$>$

Verdict: 

\end{tcolorbox}
\end{center}

\subsection{Judge Prompt}
\label{subsec:judge_prompt}

\begin{center}
\begin{tcolorbox}[width=0.95\textwidth]
Please act as an impartial judge and evaluate the quality of the responses provided by two AI assistants to the user question displayed below. Your evaluation should consider correctness and helpfulness. You will be given a reference answer, assistant A's answer, and assistant B's answer. Your job is to evaluate which assistant’s answer is better.
Begin your evaluation by comparing both assistants' answers with the reference answer. Identify and correct any mistakes. Avoid any position biases and ensure that the order in which the responses were presented does not influence your decision. Do not allow the length of the responses to influence your evaluation. Do not favor certain names of the assistants. Be as objective as possible. 
\textcolor{black}{Do not provide your explanation, just directly output your final verdict} by strictly following this format: "[[A]]" if assistant A is better, "[[B]]" if assistant B is better, and "[[C]]" for a tie.

\vspace{1em}

Question: $<$\texttt{QUESTION}$>$

Assistant A's Answer: $<$\texttt{PREDICTION\_A}$>$

Assistant B's Answer: $<$\texttt{PREDICTION\_B}$>$

Verdict: 

\end{tcolorbox}
\end{center}

\subsection{Majority Vote Prompt}
\label{subsec:majority_vote_prompt}

\begin{center}
\begin{tcolorbox}[width=0.95\textwidth]
You are a majority voter judge that decides the most common answer. 
Given a set of answers to the same problem provided by different agents, determine the answer that the majority of agents think as the correct answer. 
Do not provide any reasoning or additional text. Just the answer. 

\vspace{1em}

Question: $<$\texttt{QUESTION}$>$

Assistant 1's Answer: $<$\texttt{PREDICTION\_1}$>$

Assistant 2's Answer: $<$\texttt{PREDICTION\_2}$>$

Assistant 3's Answer: $<$\texttt{PREDICTION\_3}$>$

...

Assistant N's Answer: $<$\texttt{PREDICTION\_N}$>$

Majority Answer:

\end{tcolorbox}
\end{center}

\subsection{Rollback Prompt}
\label{subsec:rollback_prompt}

\begin{center}
\begin{tcolorbox}[width=0.95\textwidth]

You are a scorer that tells whether two answers are same. I'll provide Ground truth and Prediction. 
You have to tell that whether the prediction is saying a correct answer same as the ground truth. 
If the prediction is a sentence that include the ground truth, then it is correct answer. 
If the prediction is an equation that ends up with the ground truth, then it is correct answer. 
If the answer is correct, output your final verdict by strictly following this format: "[[Correct]]"
If the answer is incorrect, output your final verdict by strictly following this format: "[[Incorrect]]"

\vspace{1em}

Original Answer: $<$\texttt{ORIGINAL\_ANSWER}$>$

Revised Answer: $<$\texttt{REVISED\_ANSWER}$>$

Verdict: 

\end{tcolorbox}
\end{center}

\subsection{Self-Refine/Advanced-Refine Prompt}
\label{subsec:self_refine_prompt}

\begin{center}
\begin{tcolorbox}[width=0.95\textwidth]
You are a validator tasked with evaluating the quality of task results. Your job is to provide constructive feedback aimed at improving the answer. Do not provide or suggest a corrected answer—only point out what is misaligned with the given problem, any misleading reasoning, or gaps in logic or execution.
If the task involves coding, provide feedback that helps guide the generation of a working solution. This includes checking whether the syntax is correct, whether the code meets the task requirements, and pointing out potential bugs or incorrect assumptions. Again, do not write or suggest the corrected code—only critique what's wrong or missing.

\vspace{1em}

Question: $<$\texttt{QUESTION}$>$

Original Answer: $<$\texttt{ORIGINAL\_ANSWER}$>$

Your Feedback:

\end{tcolorbox}
\end{center}

\subsection{Debate - Round 2 Prompt}
\label{subsec:debate_round_2_prompt}

\begin{center}
\begin{tcolorbox}[width=0.95\textwidth]
Given the context and question, you have answered like follows. 
Check your colleagues' answer and revise your answer if necessary. 
Revise your answer without being verbose;

\vspace{1em}

Question: $<$\texttt{QUESTION}$>$

Your Answer: $<$\texttt{ORIGINAL\_ANSWER}$>$

Colleague 1's Answer: $<$\texttt{PREDICTION\_1}$>$

Colleague 2's Answer: $<$\texttt{PREDICTION\_2}$>$

... 

Colleague N's Answer: $<$\texttt{PREDICTION\_N}$>$

Revised Answer:

\end{tcolorbox}
\end{center}

\subsection{Debate - Judge Prompt}
\label{subsec:debate_final_prompt}

\begin{center}
\begin{tcolorbox}[width=0.95\textwidth]
Please act as an impartial judge and evaluate the quality of the responses provided by two AI assistants to the user question displayed below. Your evaluation should consider correctness and helpfulness. You will be given a reference answer, assistant A's answer, and assistant B's answer. Your job is to evaluate which assistant’s answer is better.
Begin your evaluation by comparing both assistants' answers with the reference answer. Identify and correct any mistakes. Avoid any position biases and ensure that the order in which the responses were presented does not influence your decision. Do not allow the length of the responses to influence your evaluation. Do not favor certain names of the assistants. Be as objective as possible. 
\textcolor{black}{Do not provide your explanation, just directly output your final verdict} by strictly following this format: "[[A]]" if assistant A is better, "[[B]]" if assistant B is better, and "[[C]]" for a tie.

\vspace{1em}

\vspace{1em}

Context: $<$\texttt{CONTEXT}$>$

Question: $<$\texttt{QUESTION}$>$

Assistant 1's Answer: $<$\texttt{PREDICTION\_1}$>$

Assistant 2's Answer: $<$\texttt{PREDICTION\_2}$>$

... 

Assistant N's Answer: $<$\texttt{PREDICTION\_N}$>$

Your Verdict:

\end{tcolorbox}
\end{center}

\subsection{LLM Planner Prompt}
\label{subsec:llm_planner_prompt}

\begin{center}
\begin{tcolorbox}[width=0.95\textwidth]
You are a workflow planner. Your task is to break down a given high-level task into an efficient and practical  workflow that  maximizes concurrency while minimizing complexity . The breakdown is meant to  improve efficiency  through  parallel execution , but  only  where meaningful. The goal is to ensure that the workflow remains  simple, scalable, and manageable  while avoiding excessive fragmentation.

--

Guidelines for Workflow Design 

\vspace{2em}
{1. Subtask Clarity and Completeness}

\begin{itemize}[leftmargin=*,nosep,topsep=1pt,parsep=2pt]
    \item Each subtask must be well-defined, self-contained, and easy to execute by a single agent.
    \item Ensure that the workflow meets all requirements of the task.
    \item Keep descriptions concise but informative.  Clearly specify the subtask's purpose, the operation it performs, and its role in the overall workflow.
    \item Avoid unnecessary subtasks.  If a task can be handled efficiently in one step without blocking others, do not split it further.
    \item Avoid repeating the same reasoning across tasks or nodes.  Solve each problem step by step, and reuse previously computed results instead of redoing reasoning.
\end{itemize}

\vspace{2em}

{2. Dependency Optimization and Parallelization}

\begin{itemize}[leftmargin=*,nosep,topsep=1pt,parsep=2pt]
    \item  Identify only necessary dependencies.  Do not introduce dependencies unless a subtask *genuinely* requires the output of another.
    \item  Encourage parallel execution, but do not force it.  If tasks can run independently without affecting quality, prioritize concurrency. However, avoid excessive parallelization that may lead to synchronization issues.
    \item  Keep the dependency graph simple.  Avoid deep dependency chains that increase complexity.
    \item  Terminal node should be only one.  There should be only one terminal leaf node. 
\end{itemize}

\vspace{2em}

3. Efficient Agent Assignment 
\begin{itemize}[leftmargin=*,nosep,topsep=1pt,parsep=2pt]
    \item Assign exactly one agent per subtask.  Every subtask must have a responsible agent.
    \item Use sequential agent IDs starting from "Agent 0".  Assign agents in a clear, structured way.
    \item Ensure logical role assignments.  Each agent should have a well-defined function relevant to the assigned subtask.
\end{itemize}

\vspace{2em}
4. Workflow Simplicity, Modularity, and Maintainability 
\begin{itemize}[leftmargin=*,nosep,topsep=1pt,parsep=2pt]
    \item Keep each subtask modular and appropriately scoped . A single node should perform a cohesive, reasonably sized operation that can be executed in one LLM call. Avoid bundling multiple distinct steps (e.g., aggregation plus external tool use plus reasoning) into a single task.
    \item Design for global simplicity.  The overall workflow should have a balanced number of subtasks—enough to promote clarity and concurrency, but not so many that it creates excessive coordination or cognitive overhead.
    \item Maintain clarity and logical flow.  The breakdown should be intuitive, avoiding redundant or trivial steps.
    \item Prioritize quality over extreme concurrency.  Do not split tasks into too many small fragments if it negatively impacts output quality.
\end{itemize}

\vspace{3em}

.. \textit{Continued on the next page.}
\end{tcolorbox}
\end{center}

\begin{center}
\begin{tcolorbox}[width=0.95\textwidth]

 5. Tool Invocation and External Knowledge Access 
\begin{itemize}[leftmargin=*,nosep,topsep=1pt,parsep=2pt]
    \item Determine if external tools are needed. If the task requires factual information, real-time knowledge, or external data, consider adding a subtask that invokes tools like web search or document retrieval.
    \item Add retrieval nodes explicitly. Create a separate node for fact-gathering with a clear objective, such as "search for the latest information on X".
    \item Link dependencies carefully. Ensure that any task using external knowledge depends explicitly on the corresponding retrieval node.
    \item Avoid blind tool use. Do not invoke tools unless the task clearly justifies it; prefer reasoning with available context if sufficient.
    \item Explicitly mention "Use tools" in the task objective for this retrieval nodes. 
\end{itemize}

\vspace{2em}

Provide the workflow in  json  format. 

\end{tcolorbox}
\end{center}


\end{document}